# Statistical investigation of the large-area Si(Li) detectors mass-produced for the GAPS experiment

M. Kozai[a,b,]*, K. Tokunaga[b], H. Fuke[b], M. Yamada[c], C.J. Hailey[d], C. Kato[e], D. Kraych[d], M. Law[d],

E. Martinez[d], K. Munakata[e], K. Perez[f], F. Rogers[f], N. Saffold[d], Y. Shimizu[g], K. Tokuda[c], M. Xiao[f]

*Corresponding author. *Email address:* kozai.masayoshi@nipr.ac.jp

[a]*Polar Environment Data Science Center, Joint Support-Center for Data Science Research, Research Organization of Information and Systems (PEDSC/ROIS-DS), Tachikawa, Tokyo 190-0014, Japan*
[b]*Institute of Space and Astronautical Science, Japan Aerospace Exploration Agency (ISAS/JAXA), Sagamihara, Kanagawa 252-5210, Japan*
[c]*Technology Research Laboratory, Shimadzu Corporation, Atsugi, Kanagawa 243-0213, Japan*
[d]*Columbia University, New York, NY 10027, USA*
[e]*Faculty of Science, Shinshu University, Matsumoto, Nagano 390-8621, Japan*
[f]*Massachusetts Institute of Technology, Cambridge, MA 02139, USA*
[g]*Kanagawa University, Yokohama, Kanagawa 221-8686, Japan*

**Abstract**

The lithium-drifted silicon (Si(Li)) detector developed for the General Antiparticle Spectrometer (GAPS) experiment features a thick (~2.2 mm) sensitive layer, large (10 cm) diameter, and excellent energy resolution (~4 keV for 20-100 keV X-rays) at a relatively high operating temperature (approximately −40°C). Mass production of GAPS Si(Li) detectors has been performed to construct a large-volume silicon tracker for GAPS. We achieved the first success of the mass production of large-area Si(Li) detectors with a high (~90%) yield rate. Valuable datasets related to detector fabrication, such as detector performance and manufacturing parameters, were recorded and collected during the mass production. This study analyzes the datasets using statistical methods with the aim of comprehensively examining the mass production and to gain valuable insight into the fabrication method. Sufficient uniformities of the performance parameters (leakage current and capacitance) between detectors and strips are found, demonstrating high-quality and stable mass production. We also search for correlations between detector performance and manufacturing parameters by using data-mining techniques. Conventional multivariate analysis (multiple regression analysis) and machine-learning techniques (regression tree analysis) are complementarily used, and it is found that the Li-drift process makes a significant contribution to the performance parameters of the finished detectors. Detailed investigation of the drift process is performed using environmental data, and physical interpretations are presented. Our results provide valuable insight into the fabrication methods for this kind of large-area Si(Li) detector, and encourages future projects that require large-volume silicon trackers.

*Keywords:*

# 1. Introduction

Cosmic-ray antinuclei have attracted attention as a key for unraveling questions in modern physics[1]. The General Antiparticle Spectrometer (GAPS) aims for high-sensitivity observation of cosmic-ray antinuclei, particularly low-energy antideuterons, which are predicted to form a distinct signal from dark matter[2–4]. The first flight using a long-duration balloon from Antarctica is scheduled for late 2022.

The novel GAPS detection concept, which is based on exotic atom physics, is achieved by using a large-volume silicon tracker with ~9,000 cm$^2$ active area and ~2.5 cm sensitive thickness in total. We developed a mass-production method for large-area (10 cm-diameter) Si(Li) detectors and produced over 1000 detectors with a high (~90%) yield rate to build the silicon tracker. It is the first success of the mass production of large-area Si(Li) detectors. Investigation of the mass-production data allows us to assess what are the main factors that influence yield.

Basic Si(Li) detector fabrication is performed by[5,6]

1. procuring a boron-doped silicon wafer,
2. evaporating/diffusing Li ions onto the wafer to form an $n^+$-layer, and
3. drifting Li ions toward the *p*-side by applying a bias voltage to the heated wafer.

The drifted Li ions compensate the boron acceptors to form a thick intrinsic layer that functions as a sensitive layer in the detector operation.

Small-diameter Si(Li) detectors[e.g., 7–11] have already been commercialized mainly in the energy dispersive X-ray (EDX) spectroscopy field. In fields such as astrophysics, detection targets are not only X-rays but also charged particles, and attempts have been made to enlarge the aperture of the Si(Li) detector[12–17]. By combining previous findings with new development, we successfully established a fabrication method for a large-area Si(Li) detector optimized for GAPS[18–21]. The main goals of development were as follows.

- ➢ Production of uniform and lightly boron-doped Si ingot, which is ideal for Si(Li) detectors. Its diameter must be as large as 10 cm.
- ➢ Uniform Li-diffusion/drift for large-diameter (10 cm) wafers.
- ➢ Noise suppression at a relatively high operation temperature (approximately −40°C), which enables the use of a low-power and lightweight cooling system[22–25].
- ➢ Establishment of a simple and high-quality mass-production method that allows for integration of a large-volume tracker consisting of many detectors.

After the manufacturing method was settled, mass production of over 1000 units was performed from 2018 to 2020 at Shimadzu Corp., in collaboration with JAXA aiming for the first flight. We conducted measurement of the detector leakage current (LC) and capacitance at room temperature (RT) immediately after fabrication of each detector, with the aim of evaluating the detector performance and as a soundness check of the production flow. The LC is a major source of noise and its suppression is effective for improving the energy resolution. It is also desirable to reduce the capacitance as much as possible, not only to suppress noise but also to ensure sufficient thickness of the sensitive layer because the capacitance is inversely proportional to the thickness of the depleted layer.

The characteristics at the operating temperature (approximately −40°C) can be estimated from the temperature dependences that were measured in the R&D phase and the initial stage of mass production. After verifying the performance based on the criteria at RT, the mass-produced detectors were transported to a series of U.S. institutes. Columbia University conducts passivation of the detector surface, and MIT and University of Hawaii at Manoa evaluate the detector characteristics including the energy resolution at operating temperatures. The development of passivation is reported in [21], and the detailed evaluation of the energy resolution of the mass-production model is reported in [20].

There have been attempts to apply data-mining methodologies to industrial fields[e.g., 20–24], to enhance production yields, and to gain valuable insight. This study presents statistical examination of the mass-produced detectors using datasets recorded during mass production, such as performance (LC and capacitance) and manufacturing parameters. The mass-produced detectors are demonstrated to have sufficiently high quality and uniformity. By using data-mining techniques, we also investigate correlations between the performance and manufacturing parameters. The findings deduced in the R&D phase[18], such as LC suppression by an undrifted layer, are verified from a new viewpoint based on statistical analysis. New insights related to the fabrication methods are also introduced.

In previous studies of large-area Si(Li) detectors, only a small number of detectors have been produced, and the degree of reproducibility was unknown in many cases. In addition, it is rare for companies to provide detailed manufacturing data to scientists. This study analyzes all of the mass-produced detectors and comprehensively handles non-uniformities and fluctuations possibly arising from manual work and the fabrication environment. Full cooperation from the manufacturing companies (Shimadzu Corp. and SUMCO Corp.) makes it possible to access a variety of the manufacturing data.

Subsection 2.1 briefly introduces the manufacturing method of GAPS Si(Li) detectors. Requirements, grading and yield rate of the detectors are described in subsections 2.2 and 2.3. Subsection 2.4 describes history of the mass production.

Section 3 investigates variations in the performance between detectors by using data-mining techniques. Multivariate analyses are performed to find correlations between the performance and manufacturing parameters. Physical interpretations for the results are provided.

Section 4 analyzes the inter-strip variation of the performance parameters and demonstrates uniformity of the performance in each detector.

Section 5 summarizes our investigations, results, and physical interpretations.

## 2. Mass production of GAPS Si(Li) detectors

2.1. Fabrication process

*Location of Figure 1.*

This section briefly describes the fabrication process of the GAPS Si(Li) detector (Fig. 1). For the detailed description, readers can refer to [18].

1) Procurement of *p*-type Si crystal

Purity and uniformity of the Si crystal are essential for obtaining a uniform and large-area Li-drifted layer[31,32]. It is also desirable to realize as light a boron doping as possible for sufficient and uniform compensation of the boron acceptors by Li ions. We developed a custom Si ingot with a diameter of ~10 cm and featuring a carrier lifetime of ~1 ms, resistivity of ~1000 Ω cm, and oxygen and carbon concentrations below the measurement limit (~$10^{16}$ atoms/cm$^3$). The floating-zone method was used for crystal growth to avoid contamination by impurities. The crystal orientation is <111>. Ingots are sliced into wafers with ~10 cm diameters and ~2.5 mm thicknesses.

2) Li evaporation and diffusion

Li is evaporated and diffused to a shallow depth to form the Li-diffused layer or $n^+$-layer. Custom Li evaporator and heater are developed to realize a uniform Li-diffused layer. Theoretical calculations and inspections of the wafer cross-section by copper-staining (see [33,34] and Fig. A3 in Appendix) confirm that the thickness of the Li-diffused layer is ~0.1 mm.

3) Evaporation of *n*-electrode and top-hat machining

Nickel and gold are evaporated on the *n*-side as an electrode with a thickness of ~140 nm. The circumference of the *n*-side is then ground by ultrasonic impact grinding (UIG) to form a top-hat geometry. This geometry prevents avalanche breakdown and inhibits Li ions from drifting to the sides of the wafer during the Li-drift process.

4) Li drift

*Location of Figure 2.*

A custom drift stage was developed to uniformly heat the large-area wafer. The wafer is put on the drift stage (hot plate in step 4 in Figure 1), which is electrically grounded, with the *p*-side facing down. A positive bias voltage $V = 600$ [V] is applied to the *n*-side by an electrode from a voltage supply. A temperature sensor (resistance temperature detector) is installed on the *n*-side of the wafer for feedback control of the heater built into the drift stage. The wafer is maintained at a constant temperature of $T = 100°C$.

Li is drifted into the bulk of the wafer by the electric field induced by the bias voltage. The drifted depth $W$ grows with drift time $t$, as [6]

$$W = \sqrt{2V\mu_L t} \qquad (1)$$

where $\mu_L$ is the Li mobility related to the diffusion constant (*D*) by the Einstein relation $\mu_L = qD/k_B T$ with the elementary charge *q* and the Boltzmann constant $k_B$. Figure 13a displays the *t-W* relationship, which is used in section 5 to discuss the interpretations of the analysis results in this study.

The output current from the bias voltage supply is recorded together with the bias voltage *V*, wafer temperature *T*, and heater output. We refer to this current as the drift current to avoid confusion with the strip LC as a performance parameter of the produced detector.

Figure 2a shows a typical drift profile (drift current and heater output). The drift current increases as the Li-drifted layer grows. The Joule heat generated by the drift current also increases and the heater output gradually decreases to compensate for the heat generation. The depleted layer containing the drifted layer expands slightly beyond the drift front and into the undrifted layer. In the final stage of the drift, the depleted layer reaches near the *p*-side surface. At this time, the drift current increases exponentially and the heater output reaches ~0%. The bias voltage supply is configured to shut down automatically when the drift current reaches 25 mA or the heater output reaches 0%, and the drift process is terminated. The bias voltage supply is also terminated manually if the operator notices an exponential increase of over 15 mA in the drift current, even if the drift current and heater output do not reach the threshold (25 mA or 0%) for automatic termination. The typical drift time is 6000-6500 min.

This sequence leaves an undrifted layer of ~0.1 mm on the *p*-side. The exponential increase of the drift current at the end of the drift is probably due to that the undrifted layer on the *p*-side becomes too thin to suppress the drift current. The undrifted layer, which is a lightly doped *p*-layer, forms a *p-n* junction with the $n^+$-layer on the *n*-side and suppresses the reverse current by the rectifying property. We guess that this effect is broken in the condition at the end of the drift, in which the undrifted layer becomes thinner and the temperature and bias voltage are much high as 100°C and 600 V respectively. On the other hand, we demonstrated that ~0.1 mm undrifted layer sufficiently works to reduce the LC at the operating temperature (approximately −40°C) [18]. Even at RT, the LC is suppressed to a reasonable level as shown in this paper.

In some cases, the drift current rapidly increases beyond the expected rate probably due to fine irregularities at the electrode edge or impurities on the wafer surface. The drift is automatically terminated in a short time (<5300 min). In these cases, the surface condition is improved by etching the top-hat side, and a second drift is performed by mounting the wafer on the drift stage again. An example of this kind of multiple drifts is displayed in Fig. 2b. Surface etching or restarting the drift sequence can reduce the drift current and enable a total drift time of >6000 min in many cases. In the case of Fig. 2b, the second drift is manually terminated so that the total drift time does not exceed 6200 min (see subsection 2.2). The maximum number of multiple drifts is three.

5) Machining grooves for the guard ring and strips

The intrinsic layer exposed on the side surface of the top-hat is relatively easy to be contaminated and can become a source or path for surface LC. A circular groove of depth ~0.3 mm and width ~1 mm is cut into the *n*-side by UIG. This groove electrically isolates the perimeter, or guard-ring[6] electrode connected to the side surface, from the central area, or readout electrode. Grooves dividing the readout electrode into 8 strips of equal area are cut at the same time as machining the guard-ring groove, using the same groove depth and width.

6) Evaporation of *p*-electrode

The metal contact on the *p*-side is evaporated in the same manner as the *n*-electrode.

7) Etching on side of the top-hat and grooves

The side of the top-hat and the *n*-side grooves are etched after painting wax on the *n*- and *p*-electrodes. This etching not only removes the damaged layer on the surface formed by UIG, but also smooths the surfaces and removes contaminants from the exposed silicon crystal. Organic-solvent cleaning is then performed to remove the wax.

This surface treatment is effective for reducing the surface LC, but unnecessarily long etching needs to be avoided because it expands the area of the grooves where the intrinsic layer is exposed. [18] deduced an optimum etching time at which the LC suppression begins to saturate. In some cases where the LC is not sufficiently suppressed by etching for the specified time, second and third etchings were additionally performed during mass production.

## 2.2. Requirements and grading of the detectors

The dataset collected in the mass production is summarized in Table 1. Excluding the 17 detectors that are missing some data other than parameters denoted by a single asterisk in Table 1, this study analyzes data from 1095 mass-produced detectors. Among them, 960 (~88%) detectors were graded as *Very Good* by the performance evaluation performed immediately after each detector was fabricated. The *Very Good* grade is defined by the LC and capacitance of each detector at RT and vacuum as follows, where RT and vacuum are defined as 18-25°C and ~$10^{-3}$ Pa, respectively.

(1) 7 strips have LCs $\leq 7$ μA at a bias voltage of 250 V (Each detector has 8 strips in total).
(2) All strips and the guard ring have LC $\leq 50$ μA at 250 V, or their rate of change with respect to the bias voltage are $\leq 0.05$ μA/V in the range from 200 to 300 V.
(3) Capacitances of all strips are in the range of 35 to 42 pF at 250 V.

*Location of Figure 3.*

Figure 3 shows a histogram of the strip LC and capacitance for all detectors and *Very Good* detectors. Modes of the distributions are around 2 μA and 39 pF for the strip LC and capacitance, respectively. These are lower than the upper limits, 7 μA and 42 pF in the criteria (1) and (3). Small part of detectors exceeds 7 μA even in the *Very Good* detectors, because the criterion (1) allows one strip in each detector to exceed 7 μA. It is also noted that most of the strip capacitances are in the range of 38 to 40 pF (~5% of the magnitude). This indicates that the variation in the thickness of the sensitive layer is within this range between strips and detectors.

This definition of the *Very Good* grade (criteria (1) to (3) above) was derived based on the following considerations. In order to identify antideuteron events from abundant antiproton events, which are the main background events, silicon detectors are required to identify characteristic X-rays from exotic atoms composed of each antinucleus species. The energy resolution necessary to meet this requirement is below ~4 keV (full width at half maximum) for 20-100 keV X-rays. The operating temperature is relatively high at approximately −40°C because we use a passive cooling system featuring low-power and lightweight cooling system [22–25]. Including a margin of +5°C, we impose a temperature of −35°C as a reference temperature for the energy resolution.

Energy resolution is easily affected by the measurement environment, takes time to measure, and does not directly reflect the detector structure such as the thickness of the sensitive layer. Thus, we adopt the LC and capacitance as primary parameters for the performance evaluation. Strip LC of below 5 nA and strip capacitance of below 40 pF were derived as requirement values to achieve the necessary energy resolution at −35°C [35]. It is also noted that the capacitance's requirement, 40 pF, also ensures a sufficient thickness of the sensitive layer.

The corresponding values at RT were derived from the temperature dependence of the detector performances, and the *Very Good* criteria (1) to (3) was defined. The temperature dependence of the strip LC is discussed in the subsection 2.3. The strip capacitance is less dependent on the temperature than the LC, but increases 1-2 pF at RT compared with −35°C. The strip LC of 7 μA in (1) and capacitance of 42 pF in (3) are the RT criteria corresponding to 5 nA and 40 pF at −35°C, respectively. Even if a strip is not capable of identifying the characteristic X-rays, it can be used to track charged particles if criterion (2) is satisfied. Therefore, the criterion (1) does not impose ≤7 μA to all strips (8 strips) in each detector. Criterion (2) is not only based on this scientific consideration, but also the demands on the power supply. The lower limit on the strip capacitance of 35 pF in criterion (3) is imposed because the capacitance cannot be lower than this value even if the detector is fully depleted.

The *Very Good* criteria at RT were defined around the initial stage of the mass production. Currently, measurements of the mass-produced detectors at cold temperatures are ongoing in the U.S. institutes. Those results may lead to some modification of the RT criteria, but will not affect our conclusions because the RT criteria are used only for ad hoc grading in this paper. RT performances measured for all detectors are used in subsequent sections. Statistical investigation of the cold temperature characteristics will be presented in future works. It is also noted that all RT performances were measured immediately after each detector was fabricated. Therefore, it is advantageous to use the RT data so that we do not need to consider extra effects, such as detector storage or shipping.

Requirements of the dimensions are that the overall diameter is within 98.4-100.4 mm, the guard-ring outer diameter is within 95.5-96.8 mm, the diameter of the sensitive area is within 89.0-90.0 mm, and the overall thickness is within 2.35-2.53 mm. Definitions of these parameters are described in Table 1. All of the mass-produced detectors meet these requirements.

## 2.3. Temperature dependence of the strip LCs

*Location of Figure 4.*

In the initial stage of the mass production, the cold temperature (−35°C) performance was measured in addition to the RT performance for several detectors in order to examine the temperature dependence of the mass-produced detectors. We briefly demonstrate the temperature dependence using the strip LCs of these mass-produced detectors (5 detectors, i.e., 40 readout strips) measured at the cold temperature. Each point in Fig. 4 corresponds to a strip. The horizontal axis is the strip LC at RT, and the vertical axis is the strip LC at −35°C. The red vertical line is an upper limit (7 μA) at RT defined in the criterion (1), while the blue horizontal line is the upper limit (5 nA) at −35°C. Most of the strips are distributed around a straight line, y/x = 1/7000, in the log-log plot, indicating the uniformity of the temperature dependence of the strip LC. However, note that there is an outlier for which the ratio of the y-value to the x-value is ~1/2000. Even if we adopt this temperature dependence (y/x ≈ 1/2000), the upper limit (7 μA) at RT corresponds to 3-4 nA at −35°C which still meets the requirement (5 nA) at −35°C.

## 2.4. History of the mass production

*Location of Figure 5.*

At the last of the section 2, we briefly describe the history of the mass production. Figure 5 shows the cumulative number of detectors for the entire period of the mass production. The black curve shows the total number of detectors and the blue curve shows the number of detectors graded as *Very Good*. In January, May, and August of 2019 and January of 2020, the cumulative number temporarily stops increasing due to national holidays.

The figure inside Fig. 5 shows an enlargement of ~3.5 months in the early stage of the mass production. In the period until the middle of February 2019 (period A) indicated by the red arrows, the yield rate of *Very Good* detectors was only ~50%. In this period, only the increase (exceeding 25 mA or an exponential increase) of the drift current or the elimination (0%) of the heater output was imposed as a condition for terminating Li-drift, as described in subsection 2.1. This automatic termination was effective for ensuring an optimal thickness of the undrifted layer for the testing detectors in the R&D phase. However, in period A, relatively-low increasing rates of the drift currents, resulting in extremely long (>6500 min) drift times, were observed mainly in the detectors with poor performances.

We imposed a limit on the drift time from the middle of February 2019 to control this phenomenon, which was unexpected before the start of mass production. The thickness of the detector is 2.5 mm with an error below 0.05 mm. Since the $n^+$-layer thickness is ~0.1 mm, the drifted layer needs to be ≲2.3 mm to ensure a sufficient thickness (~0.1 mm) of the undrifted layer. In the case where the drift time is 6200 min, the drifted layer is ~2.2 mm based on equation (1) and the requirement ≲2.3 mm is met with some room. The capacitance is ~38 pF/strip and the *Very Good* criterion (3) is also met. Based on this consideration, we imposed a time limit of 6200 min on the drift time from the middle of February 2019. After this modification of the drift process, performance uniformity between the fabricated detectors was improved, and the mass production of >1000 detectors was completed with a high yield rate.

It had been unclear what caused the variation in the drift parameters (drift time and drift current) between the detectors. The variation in the surrounding environment (temperature and humidity) in the drift room during the drift process is one possible cause. Temperature and humidity monitors were installed in the room from February 2019 based on this consideration. In section 3, the correlation between the drift parameters and the surrounding environment in the drift room is investigated using the data recorded during mass production. The results show that the drift current in period A is possibly suppressed due to relatively low humidity, causing a long drift time.

**Table 1:** List of datasets of mass-produced detectors collected in this study.

1. Performance parameters

| | |
|---|---|
| Strip LC [A] | Measured at a bias voltage of 250 V. |
| **Slope of the strip LC [A/V] | Change rate of the strip LC between bias voltages of 200 V and 300 V. This parameter is not used in data analyses in this study because it shows a high correlation coefficient (r ~ 0.9) with the strip LC. |
| Strip capacitance [pF] | Measured at a bias voltage of 250 V. |

2. Measurement parameter

| | |
|---|---|
| Measurement temperature [°C] | The RT when the performance parameters were measured. |

3. Manufacturing parameters

| | |
|---|---|
| Career lifetime [μsec] | This was measured for each Si ingot, which corresponds to ~45 detectors. |
| Number of drifts | Multiple drifts were performed for detectors in which the first drift was terminated early (see item 4 in subsection 2.1). This parameter is the total number of drifts in each detector. |
| Drift time [min] | In this study, the drift time is defined as the length of time when the bias voltage of 600 V is applied in the Li-drift. In the case of multiple drifts, drift times are summed up for all drifts in each detector, as the "total drift time" $t_{\text{drift}}$ for use in the data analyses. |
| Drift current [mA] | Time variation in the drift current in each drift was recorded, but it is averaged over all drift time in each detector, as the "average drift current" $I_{\text{drift}}$, for use in the data analyses. |
| **Heater output [%] during Li-drift | Time variation in the heater output in each drift was recorded in arbitrary unit [%] together with the drift current. This parameter is not used in the data analyses because it shows a large correlation coefficient (approximately −0.8) with the drift current due to the feedback control (see item 4 in subsection 2.1). |
| *Drift room temperature [°C] | RT of the drift room (the room for the Li-drift process). This parameter is averaged over all drift time in each detector, as the "average drift temperature" $T_{\text{drift}}$, for use in the data analyses. |
| *Drift room humidity [%r.h.] | Relative humidity of the drift room. This parameter is averaged over all drift time in each detector, as the "average drift humidity" $H_{\text{drift}}$ in this study. |
| Number of etchings in the final process | Multiple etchings on the top-hat side and isolation grooves were performed in the case where a sufficiently low LC was not recorded after the first etching (see item 7 in subsection 2.1). This parameter is the total number of these etchings in each detector. |

*: Measured only for the detectors whose Li-drift processes were completed after the end of February 2019.

**: Not used in the data analyses in this paper due to a high correlation coefficient with another parameter.

4. Dimension parameters [mm]

| Overall diameter | Overall diameter of the detector. |
|---|---|
| Guard-ring outer diameter | Diameter of the guard-ring outer edge. |
| Diameter of the sensitive area | Diameter of the circle area consisting of 8 readout strips. |
| Width of the guard ring | Width of the guard-ring electrode. |
| Overall thickness | Overall thickness of the detector. |

# 3. Investigation of the relation between the detector performance and manufacturing parameters

## 3.1. Data processing

In this section, we use data-mining techniques to investigate which factors (such as manufacturing parameters) affect the detector performance and how. The performance parameters measured in each strip are averaged for 8 strips in each detector by the methods introduced below. We use these representative values of each detector to investigate the performance variation between detectors in the subsequent subsections.

[18] showed that the undrifted layer left on the *p*-side is effective for suppressing LC, and that LC increases by a few orders of magnitude when the undrifted layer is removed. Optimization of the surface etching is also found to improve LC by a few orders of magnitude. Generally, LC is very sensitive to the structure near the electrodes, such as the undrifted layer, and to the surface condition. It is natural to suppose, therefore, that the strip LC has a distribution similar to a log-normal distribution.

On the other hand, the capacitance is proportional to the electrode area and inversely proportional to the thickness of the depleted layer. As described in item 5 in subsection 2.1, each strip is segmented with an equal area. Even in an extreme case that strips couple electrically with each other due to incomplete isolations of grooves, expansion of the electrode area is less than eight times, and strip capacitances do not increase by orders of magnitude. The variation range of the strip capacitance is smaller than the LC, and is supposed to have a distribution similar to a normal distribution.

Based on these considerations, the strip LC $I_{i,j}$ [A] of detector $i$ ($i = 1, ... ,1095$) and strip $j$ ($j = 1, ... , 8$) is converted to a logarithmic value $\log_{10} I_{i,j}$. The representative value of the strip LC in each detector is derived by the 8-strip log-mean,

$$\langle \log_{10} I \rangle_i = \left(\frac{1}{8} \sum_{j=1}^{8} \log_{10} I_{i,j}\right). \qquad (2)$$

The strip capacitance $C_{i,j}$ is simply averaged (8-strip mean) as

$$\langle C \rangle_i = \left(\frac{1}{8} \sum_{j=1}^{8} C_{i,j}\right). \qquad (3)$$

We show that these 8-strip log-mean and mean are appropriate as representative values for each detector in Appendix. Their histograms are shown in Fig. 6. We can see a similar trend in each histogram of all strips in Fig. 4.

*Location of Figure 6.*

## 3.2. Multiple regression analysis for the performance parameters

We perform a multiple regression analysis using the dataset in Table 1. The performance parameter ($\langle \log_{10} I \rangle_i$ or $\langle C \rangle_i$) is used as an objective variable and other parameters (measurement temperature, manufacturing parameters, and dimension parameters) are used as explanatory variables. The slope of the strip LC in Table 1 is excluded from data analyses in this study because it shows a high correlation coefficient (r ~

0.9) with the strip LC. Feedback control of the heater output that keeps the wafer temperature at 100°C ensures a large correlation coefficient (approximately −0.8) with the drift current. Therefore, the heater output is also excluded from data analyses in this study.

The parameters denoted by a single asterisk in Table 1, which are recorded only after February 2019, are excluded in this analysis in subsections 3.2 and 3.3, to maximize the number of records in the dataset. Their contributions are studied separately in subsection 3.4.

The multiple regression analyses are performed after normalizing all parameters to have average values of 0 and dispersions of 1. Tables 2 (1) and (2) show the results of the multiple regression analyses. The total drift time ($t_{\text{drift}}$) and the average drift current ($I_{\text{drift}}$) have small $p$-values ($\leq$1e−07) and significant absolute values of regression coefficients (>0.15) for both LC and capacitance. These contributions are investigated in more detail in subsection 3.3. The other parameters do not show contributions as significant as $t_{\text{drift}}$ and $I_{\text{drift}}$, as described below.

The measurement temperature (RT) does not show a significant contribution to the performance parameter, prompting the search for contributions from other parameters (manufacturing and dimension parameters).

The carrier lifetime shows a contribution to the capacitance. A large carrier lifetime indicates a small concentration of trapping centers (or impurities) that disturb the Li-drift, and allows a thick depleted layer to form. Therefore, the negative regression coefficient of the carrier lifetime for the capacitance is qualitatively reasonable. A negative contribution to the LC is also expected, but Table 2 does not show a significant contribution of the carrier lifetime to the LC. The carrier lifetime is measured for each Si ingot corresponding to ~45 detectors, unlike other parameters that are measured for each detector. More careful handling of this parameter may be necessary to find a contribution to the LC, while such an analysis is out of scope of this study.

The number of drifts also contributes only to the capacitance. This parameter is expected to indirectly reflect the drift profiles that are eliminated by taking averages, $t_{\text{drift}}$ and $I_{\text{drift}}$. For example, the beginning stage (~150 min) of each drift in which the bias voltage is ramped up from 0 to 600 V is eliminated when calculating $t_{\text{drift}}$ and $I_{\text{drift}}$. The correlation between the number of drifts and the capacitance possibly represents such an aspect. Analyses of this kind of time evolution of the drift parameters will be performed in future works.

The number of etchings does not show a contribution to the performance parameters in Table 2. Multiple etchings were performed until suppression of the LC saturates (see item 7 in subsection 2.1). In other words, the number of etchings is controlled so that the performance becomes as uniform among detectors as possible. Therefore, this insignificant contribution is reasonable.

Dimension parameters are expected to have less contributions than manufacturing parameters, because they are too uniform to arise a performance variation. However, Table 2 shows significant contributions from two dimension-parameters (guard-ring outer diameter and width of the guard-ring electrode) only on the LC. One possible story is that the surface smoothness or edge sharpness in the UIG process may correlate with the dimension variation and also affect the surface LC or internal electric field of the detector. The final etchings are effective to make a surface condition uniform, as described above. However, there may be a kind of non-uniformity that is caused by the UIG and cannot be removed by the etchings. Further investigation focusing on the machining accuracy is necessary to validate this hypothesis.

**Table 2:** Results of the multiple regression analyses. The partial regression coefficient and *t*-test results (*t*- and *p*-values) for null hypothesis are listed for each explanatory variable. The objective variable is (1) 8-strip log-mean of the LCs ($\langle \log_{10} I \rangle_i$), (2) 8-strip mean of the capacitances ($\langle C \rangle_i$), or (3) average drift current $I_{\mathrm{drift}}$, respectively for each multiple regression analysis.

| | | (1) Multiple regression analysis for 8-strip log-mean of LCs | | | (2) Multiple regression analysis for 8-strip mean of capacitances | | |
|---|---|---|---|---|---|---|---|
| | | regression coef. | *t*-value | *p*-value | regression coef. | *t*-value | *p*-value |
| Measurement temperature (RT) | | $-0.059 \pm 0.031$ | $-1.92$ | 0.055 | $0.084 \pm 0.031$ | 2.71 | *0.007 |
| Manufacturing | Carrier lifetime | $0.087 \pm 0.033$ | 2.64 | *0.009 | $-0.180 \pm 0.034$ | $-5.40$ | **8e−08 |
| | Number of drifts | $-0.013 \pm 0.034$ | $-0.39$ | 0.694 | $-0.195 \pm 0.034$ | $-5.74$ | **1e−08 |
| | Total drift time $t_{\mathrm{drift}}$ | $0.244 \pm 0.034$ | 7.11 | **2e−12 | $-0.273 \pm 0.035$ | $-7.81$ | **1e−14 |
| | Average drift current $I_{\mathrm{drift}}$ | $0.168 \pm 0.032$ | 5.30 | **1e−07 | $0.259 \pm 0.033$ | 8.03 | **3e−15 |
| | Number of etchings | $0.084 \pm 0.031$ | 2.76 | *0.006 | $-0.021 \pm 0.032$ | $-0.67$ | 0.501 |
| Dimension | Overall diameter | $-0.003 \pm 0.037$ | $-0.08$ | 0.937 | $-0.041 \pm 0.037$ | $-1.12$ | 0.262 |
| | Guard-ring outer diameter | $-0.286 \pm 0.043$ | $-6.68$ | **4e−11 | $0.042 \pm 0.043$ | 0.96 | 0.336 |
| | Diameter of the sensitive area | $-0.067 \pm 0.035$ | $-1.81$ | 0.054 | $-0.089 \pm 0.036$ | $-2.53$ | 0.012 |
| | Width of the guard-ring electrode | $0.242 \pm 0.041$ | 5.90 | **5e−09 | $-0.004 \pm 0.042$ | $-0.09$ | 0.931 |
| | Overall thickness | $-0.034 \pm 0.028$ | $-1.21$ | 0.226 | $-0.091 \pm 0.029$ | $-3.20$ | *0.001 |

| | (3) Multiple regression analysis for average drift current $I_{\mathrm{drift}}$ | | |
|---|---|---|---|
| | regression coef. | *t*-value | *p*-value |
| Average drift temperature $T_{\mathrm{drift}}$ | $0.019 \pm 0.033$ | 0.59 | 0.56 |
| Average drift humidity $H_{\mathrm{drift}}$ | $0.246 \pm 0.033$ | 7.46 | **2e−13 |

\* *p*-value <0.01; ** *p*-value <0.001

## 3.3. Regression tree analyses for performance parameters

*Location of Figure 7 and 8.*

The most significant contributions of the total drift time $t_{\text{drift}}$ and average drift current $I_{\text{drift}}$ to the performance parameters were derived in subsection 3.2. In this subsection, we investigate these contributions in detail by using regression tree analyses (rpart; R Ver 4.0.2 [36]). Figure 7 shows the distributions of $t_{\text{drift}}$ and $I_{\text{drift}}$. We use these parameters as explanatory variables in the regression tree analyses in this subsection, and use the performance parameters (LC and capacitance) as the objective variables.

Figure 8 (a) shows the results of regression tree analysis for the 8-strip log-mean of LCs ($\langle \log_{10} I \rangle_i$). The detectors with $t_{\text{drift}} \geq 6430$ min have obviously high LCs. Their sample number is $n = 58$. In the detectors with $t_{\text{drift}} < 6430$ min, 21 detectors with $I_{\text{drift}} \geq 9.33$ mA show relatively high LCs and another 1016 detectors have the lowest LCs.

Figure 8 (b) shows a regression tree for the 8-strip mean of capacitances ($\langle C \rangle_i$). The majority of the detectors (1034 detectors) with $t_{\text{drift}} \geq 5991$ min have the lowest capacitances. In the detectors with $t_{\text{drift}} < 5991$ min, 9 detectors with $I_{\text{drift}} \geq 9.40$ mA have the highest capacitances and another 54 detectors have relatively high capacitances.

Both these results demonstrate that the majority of the detectors ($n \approx 1000$) are classified in a group with the best performances (low LC and low capacitance) by regression tree learning on the drift parameters. It is also noted that the accuracies of the discrimination values, $t_{\text{drift}} = 5991$ and 6430 min and $I_{\text{drift}} = 9.33$ and 9.40 mA, represent only those of the measurements. Significant digits for the physical interpretations or discussions, on the other hand, are likely poorer by one or two orders of magnitude (~100 min and ~1 mA) than measurement accuracies, considering the non-uniformities arising from manual work or fabrication environments.

*Location of Figure 9.*

Figure 9 classifies the detectors by merging the discrimination values in the two regression trees. The detectors are classified into 3 groups (a-c) according to the total drift time. Groups (a) and (b) are also classified into 2 groups according to the average drift current. Two discrimination values (9.33 and 9.40 mA) of $I_{\text{drift}}$ in two regression trees are unified into 9.40 mA, because their difference is within the significant digits mentioned above.

The median of each histogram is represented by a red vertical line, while the first and third quartiles are shown by dotted vertical lines. The median values are summarized in Table 3 along with the number of detectors in each group. The majority of the detectors (976 detectors, group (b)) are in the range of $5991 \leq t_{\text{drift}} < 6430$ min.

The majority (964 detectors) of group (b) with $I_{\text{drift}} < 9.40$ mA have relatively low medians in both LC and capacitance, showing comprehensively the best performance. The smaller part of group (b) with $I_{\text{drift}} \geq 9.40$ mA has relatively high LC, while the capacitance is comparable to the part with $I_{\text{drift}} < 9.40$ mA.

In group (a) with $t_{drift} < 5991$ min, 54 detectors with $I_{drift} < 9.40$ mA have the lowest LC together with the those with $I_{drift} < 9.40$ mA in group (b). However, the capacitance is higher than groups (b) and (c). The 7 detectors with $I_{drift} \geq 9.40$ mA have significantly high capacitances and relatively high LCs.

The group (c) with $t_{drift} \geq 6430$ min (58 detectors) shows the lowest capacitance. However, the LC distribution is broad containing significantly high LCs.

The longer (shorter) drift time generally results in a thicker (thinner) drifted layer and thinner (thicker) undrifted layer on the *p*-side. The capacitance is inversely proportional to the thickness of the depleted layer which is approximately equal to that of drifted layer. Therefore, relatively high capacitance in group (a) and low capacitance in group (c) are reasonable. It is also proven that insufficiently thin undrifted layer causes high LC[18]. This is considered as the reason that group (c) has higher LC.

Detailed analyses and discussions related to the drift current are given in subsections 3.4 and 3.5.

**Table 3:** Summary of Fig.9. The sample number and median value of each histogram are listed.

| Groups | | Number (n) of detectors | Median values of histograms | |
|---|---|---|---|---|
| | | | 8-strip log-mean of LCs [$\log_{10}(A)$] | 8-strip mean of capacitances [pF] |
| (a) $t_{drift} \geq 5991$ min | $I_{drift} < 9.40$ mA | 54 | -5.8 | 39.3 |
| | $I_{drift} \geq 9.40$ mA | 7 | -5.5 | 41.8 |
| (b) $5991 \leq t_{drift} < 6430$ min | $I_{drift} < 9.40$ mA | 964 | -5.8 | 38.9 |
| | $I_{drift} \geq 9.40$ mA | 12 | -5.4 | 39.0 |
| (c) $t_{drift} \geq 6430$ min | | 58 | -5.2 | 38.8 |

## 3.4. Multivariate analyses for the average drift current

In addition to the total drift time $t_{\text{drift}}$, correlations between the average drift current $I_{\text{drift}}$ and the detector performances were found in subsection 3.3. The cause of the variation in the drift current between detectors is not clear, but it is reasonable to ascribe it to the variation of drift environment because the surface current is expected to be affected by the surrounding environment. In this section, we investigate the relationship between the drift room environment and average drift current $I_{\text{drift}}$.

The temperature and relative humidity in the drift room were recorded after February 2019, as mentioned in subsection 2.2. These are averaged over the drift time in each detector, as the average drift temperature $T_{\text{drift}}$ and average drift humidity $H_{\text{drift}}$ for use in the data analyses. The drift room was air-conditioned and dehumidified, but $T_{\text{drift}}$ and $H_{\text{drift}}$ vary between detector productions in the ranges of 26-32°C and 20-35%r.h. respectively.

Table 2 (3) shows the results of a multiple regression analysis using $T_{\text{drift}}$ and $H_{\text{drift}}$ as explanatory variables and $I_{\text{drift}}$ as an objective variable. All variables are normalized to have means of 0 and deviations of 1, like the multiple regression analysis in Table 1 (1) and (2). While we cannot find a contribution from $T_{\text{drift}}$, $H_{\text{drift}}$ shows a significant contribution to $I_{\text{drift}}$ with large (~0.25) regression coefficient and small (2e-13) $p$-value.

*Location of Figure 10.*

Figure 10 shows the results of the regression tree analysis for $I_{\text{drift}}$ using $H_{\text{drift}}$ as an explanatory variable ($T_{\text{drift}}$ is not used based on the multiple regression analysis). The majority of the detectors (n = 642) with $H_{\text{drift}} < 30.4$%r.h. had $I_{\text{drift}}$ less than ~5 mA, while the remainder (n = 339) with $H_{\text{drift}} \geq 30.4$%r.h. shows a broader distribution with higher $I_{\text{drift}}$. This result possibly indicates that higher humidity causes higher drift current.

## 3.5. Summary of the correlation between drift profile and detector performance

The statistical investigations in this study ensure significantly close correlation between the detector performance and the drift profile, which is represented by $t_{\text{drift}}$ and $I_{\text{drift}}$. In this subsection, we summarize the analysis results related to the drift parameters and detector performance, giving interpretations of the current findings.

If the drift room humidity is low and the drift current is suppressed, the drift current takes time to meet the threshold (>25 mA) for drift termination. In the case where no limit is imposed on the drift time and the drift time becomes too long (over ~6400 min), the strip LC increases drastically as shown in Fig. 9. This is considered to be because sufficient thickness of the undrifted layer does not remain on the *p*-side. The negative correlation between the strip capacitance and the total drift time supports this interpretation.

In the early stage of the mass-production before February 2019, the drift room environment was not recorded. However, the humidity is expected to be relatively low during the winter season in Japan. The limit on the drift time was not imposed in addition, resulting in extremely long drift times and poor yield rate of *Very Good* detectors during that period.

The strip LC is increased in detectors with the average drift current $I_{\text{drift}}$ exceeding ~9 mA, as shown in Fig. 9. The strip capacitance also has a significant correlation with $I_{\text{drift}}$. However, this occurs only in the cases that $t_{\text{drift}}$ is below ~6000 min, unlike the case of the LC. This difference between LC and capacitance possibly arises from the fact that LC is much more sensitive to the surface or internal condition of the detector.

Most of the detectors with an average drift current $I_{\text{drift}}$ exceeding ~9 mA are produced when the drift room humidity $H_{\text{drift}}$ was as high as above ~30%r.h. A high humidity is expected to change the surface state of the detector by promoting adsorption of contaminants [37]. This possibly generate current sources or paths that increase the drift current.

However, the high strip LC occurring in the cases with high $I_{\text{drift}}$ and $H_{\text{drift}}$ is not directly ascribed to contaminants from the drift room atmosphere. The reason is that surface etching is performed in the final process (item 7 in subsection 2.1) to remove the surface contaminants. One possible explanation is that disturbances in temperature or electric field [37] inside the wafer due to increased drift current or charged contaminants inhibit uniform Li-drift, and affect the internal structure of the produced detector. Another possibility is that when the drift room humidity is high, the humidity elsewhere also tends to be high and affects the produced detector surface.

## 4. Inter-strip uniformity of the strip LCs and capacitances in each detector

Inter-strip variations of the strip LCs and capacitances in each detector are investigated in this section. The results enable evaluation of the uniformity of processes such as Li-diffusion, Li-drift, and segmentation in the large-area detector.

### 4.1. Strip LCs

*Location of Figure 11.*

Square root of the variance of the logarithmic strip LCs ($\log_{10} I_{i,j}$) in each detector $i$,

$$\sigma_i(\log_{10} I) = \sqrt{\frac{\sum_{j=1}^{8}(\log_{10} I_{i,j} - \langle \log_{10} I \rangle_i)^2}{8}} \quad (4)$$

is defined as 8-strip log-deviation of the LCs. We use this parameter as an indicator of the inter-strip variation of the strip LCs. Figure 11(a) shows a scatter plot between the 8-strip log-mean ($\langle \log_{10} I \rangle_i$) and log-deviation ($\sigma_i(\log_{10} I)$) of LCs. In a minority of the detectors with $\langle \log_{10} I \rangle_i$ exceeding ~5.5 [$\log_{10}$(A)], there is likely a positive correlation between $\langle \log_{10} I \rangle_i$ and $\sigma_i(\log_{10} I)$. Figure 11 (b) shows a histogram of $\sigma_i(\log_{10} I)$. The majority of the detectors are distributed below $\sigma_i(\log_{10} I) \approx 0.2$ [$\log_{10}$(A)]. On the other hand, the 8-strip log-mean $\langle \log_{10} I \rangle_i$ distributes from approximately $-6.5$ to $-5.5$ [$\log_{10}$(A)], i.e., in a range of ~1 [$\log_{10}$(A)] as shown in Fig. 6. The 8-strip log-deviation $\sigma_i(\log_{10} I) \approx 0.2$ is ~20% of this range.

The relative deviation of the absolute value, $I_{i,j}$ of the strip LC can be estimated as

$$\frac{10^{\{\langle \log_{10} I \rangle_i + \sigma_i(\log_{10} I)\}} - 10^{\langle \log_{10} I \rangle_i}}{10^{\langle \log_{10} I \rangle_i}} = 10^{\sigma_i(\log_{10} I)} - 1. \quad (5)$$

The upper horizontal axis in Fig. 11 (b) displays this relative deviation. It is less than 0.6 (60%) in the majority of the detectors, and is 0.2 (20%) at the mode of the histogram. Only small number of the detectors have the relative deviations exceeding 1 (100%).

## 4.2. Strip capacitances

*Location of Figure 12.*

Inter-strip variation of the capacitances in each detector is investigated as well as the strip LCs. We use the square root of the variance of the strip capacitance $C_{i,j}$ in each detector,

$$\sigma_i(C) = \sqrt{\frac{\sum_{j=1}^{8}(C_{i,j} - \langle C \rangle_i)^2}{8}} \tag{6}$$

as an index of the inter-strip variation.

Figure 12(a) shows a scatter plot between the 8-strip mean ($\langle C \rangle_i$) and deviation ($\sigma_i(C)$) of the capacitance in each detector. We cannot find any systematic correlation between these parameters. This is probably due to the small (~2 pF) variation of the 8-strip mean between detectors (see the right panel of Fig. 6). Figure 12 (b) shows a histogram of the 8-strip deviation. The majority of the detectors had an 8-strip deviation of below ~0.3 pF. This is less than 1% of the average strip capacitance, ~39 pF, demonstrating excellent uniformity of the thickness of the drifted layer in each detector. It is also ~15% of the variation (~2 pF) between the detectors.

The measured overall thickness of each detector is sufficiently uniform between detectors, with a variation as small as $\pm 0.01$ mm (0.4% of the average thickness, 2.5 mm). The inter-strip variation of the overall thickness in each detector was not measured, but is expected to be of the same order as the variation between detectors. Therefore, the high inter-strip uniformity of the capacitance, or the drifted layer thickness, in each detector indicates the high inter-strip uniformity of the undrifted layer thickness on the *p*-side. The inter-strip variation of LC is, on the other hand, as large as ~20%. This is probably due to the high sensitivity of LC to slight variations in the undrifted layer thickness or surface conditions.

# 5. Summary and discussions

This study performed statistical investigation of large-area Si(Li) detectors mass-produced for the GAPS experiment. We demonstrated the quality of the produced detectors and found new insights into the fabrication methods.

The mass production was carried out to construct the GAPS silicon tracker. Stable mass-production proceeded as planned, and ~90% of the detectors passed the *Very Good* criteria imposed by the science requirements. It is the first success of the mass production of the large-area Si(Li) detectors. The strip LC is distributed around ~2 µA at RT, which is several times lower than the criteria, 7 µA. This performance margin ensures the reliability of the tracker. The inter-strip deviation of the strip capacitances is below ~1%. The variation of the capacitance between detectors is also as small as ~5% (~38 to 40 pF). This guarantees uniformity of the thickness of the depleted layer. Obtaining a uniform Li-drifted layer has been one of the main development challenges in the previous studies on large-area Si(Li) detectors [13,14,16].

LC is much more sensitive than the capacitance to the surface or internal condition of the detector, making it challenging to secure a uniformity of LC between strips or detectors. We demonstrated that inter-strip deviation of LC is below ~20% (~0.2 [$\log_{10}(A)$]) of the variation (from approximately $-6.5$ to $-5.5$ [$\log_{10}(A)$]) between detectors, as far as the logarithmic scale is concerned. Detailed evaluation of the uniformity of the LC will be done in future work investigating the total performance of the tracker, while the sufficiently low LC ensures successful operation of the tracker.

If the surface current is dominant in the LC due to poor surface conditions, electrical isolation between the strips also becomes incomplete, and the capacitance increases along with the LC. On the other hand, if the bulk current is dominant, the LC is increased when a sufficient thickness of the undrifted layer is not left near the *p*-side [18]. At the same time, the capacitance is suppressed when the undrifted layer becomes thinner. The result of the multivariate analyses in subsection 3.3 supports the latter description, showing that the higher LC, lower capacitance, and longer drift time correlate with each other.

The analysis results also demonstrate that the optimal value of the total drift time is in a range of 6000 to 6400 min. The upper limit on the drift time imposed after February 2019 in the mass production, 6200 min, is at the center of this range and proved to be a reasonable value.

*Location of Figure 13.*

This upper limit on the drift time is derived from equation (1) in the mass production to leave a thin but sufficient undrifted layer on the *p*-side. The black line in Fig. 13 (a) displays the *t-W* curve derived from equation (1). Red vertical lines indicate the range of optimal drift time (from 5991 to 6430 min) derived from the regression tree analyses. The threshold of the drifted depth where the LC starts increasing rapidly is ~2.23 mm (the drift time of ~6400 min). The detector thickness is 2.5 mm and the thickness of the $n^+$-layer is estimated to be ~0.1 mm from the copper-staining test on the cross-section. Therefore, the required thickness of the undrifted layer is estimated to be 0.1-0.2 mm. This is roughly consistent with the required thickness of ~0.1 mm obtained by *p*-side polishing in [18].

In the R&D phase of the GAPS Si(Li) detector, the effect of the undrifted layer on suppressing LC was studied by polishing the undrifted layer on *p*-side. This study proves the effect from different aspects, i.e., variations in the manufacturing and performance parameters between mass-produced detectors, by using data-mining techniques.

Figure 13 (b) shows an enlargement of the range between the two red vertical lines. The majority of the mass-produced detectors (~1000 detectors) are contained in this range. The corresponding difference in the drifted depth (difference between the blue horizontal lines) is ~0.1 mm or less, that is, ~5% of the thickness (~2.2 mm) of the drifted layer. This variation probably causes the variation (~5%) of the capacitance between detectors.

We also found a correlation between the drift current and performance parameters. The average drift current becomes high when the relative humidity of the drift room is high. In this case, the LC also increases as an effect other than the drift time. The reason is not clear but we can make the guesses shown in subsection 3.5. However, in the case where the drift current is extremely suppressed likely due to the low humidity, the threshold of the drift current or heater output (25 mA or 0%) does not work to terminate the drift process in a reasonable time. This case can be avoided by imposing an upper limit (~6200 min) on the drift time, as we did after February 2019 during the mass production.

It is demonstrated that the Li-drift is a critical process for the Si(Li) detector performance. The environmental humidity appears to be a control parameter determining the drift profile. In future fabrication, more accurate control of the humidity in the process will possibly provide more stable performances of the detectors. The drift time also specifies the detector performance. Appropriate drift time is ~6200 min for the GAPS Si(Li) detector (2.5 mm thickness), but depending on the scientific requirements and operating environment, this time may be adjusted in the case of other experiments.

Our unprecedented mass production of the large-area Si(Li) detectors and results of the statistical investigations encourage future projects that require large-volume silicon tracker using large-area Si(Li) detectors.

# APPENDIX: Distribution profiles of the strip LC and capacitance normalized in each detector

The performance parameters (strip LCs and capacitances) contain both the inter-strip variations and variations between detectors, simultaneously. In this appendix, we specifically analyze the inter-strip variation by normalizing each parameter in 8-strips in each detector to its mean, to eliminate the variation between detectors. In sections 3 and 4, we used the 8-strip log-mean and mean of the LCs and capacitance, respectively, as representative values of each detector. The validity of this calculation is discussed by using the normalized parameters. Performance dependence on the strip position (strips A-H in Fig. 1) is also investigated.

## A1. Data processing

The performance parameter $x_{i,j}$ of strip $i$ in detector $j$ is normalized as

$$z_{i,j} = \frac{x_{i,j} - \langle x \rangle_i}{\sigma_i} \quad (i = 1 \ldots 1095, j = 1 \ldots 8) \tag{A1}$$

$$\text{where } \langle x \rangle_i = \frac{1}{8}\sum_{j=1}^{8} x_{i,j} \text{ and } \sigma_i = \sqrt{\frac{\sum_{j=1}^{8}(x_{i,j} - \langle x \rangle_i)^2}{8}}. \tag{A2}$$

The mean value of the eight $z_{i,j}$s in each detector is 0, and its deviation is 1. The strip LC (or its logarithmic value) and capacitance are adopted as $x_{i,j}$ in the following analyses. Variations between detectors are eliminated by this normalization, allowing inter-strip distributions to be extracted for each detector.

## A2. Strip LCs

### A2.1. Distribution profile in each detector

The dotted line (solid line) histogram in Fig. A1 (a) shows a histogram of $z_{i,j}$ calculated by using the absolute value (logarithmic value) of the strip LC $I_{i,j}$ [A] as $x_{i,j}$. This histogram represents the average profile of the inter-strip distribution in each detector. Numbers (*n* or n) of the strips with $z_{i,j} < 0$ and $z_{i,j} \geq 0$ are displayed in the histogram. The italic character *n* denotes the number of the dotted line histogram, while the normal character n denotes the number of the solid line histogram. The dotted line histogram is biased to negative $z_{i,j}$ with a peak at $z_{i,j} \approx -0.5$. This indicates that a certain number of large outliers is contained in the 8 strip LCs in many detectors. On the other hand, the solid line histogram has comparable numbers of samples in $z_{i,j} < 0$ and $z_{i,j} \geq 0$ compared to the dotted line histogram. That is, the bias seen in the absolute LC distribution is eliminated by the logarithmic conversion.

*Location of Figure A1.*

Figure A1 (a) and the related discussions above still have an uncertainty, because it reflects not only the inter-strip distribution in each detector but also the variation in the distribution profile between detectors. We calculate the median of 8 values of $z_{i,j}$ in each detector to extract a parameter representing the distribution profile in each detector. The median $[z]_i$ is derived as

$$[z]_i = \frac{1}{2}(z'_{i,4} + z'_{i,5}) \tag{A3}$$

where $z'_{i,1}, z'_{i,2}, \ldots, z'_{i,8}$ are the same as $z_{i,j}$ but the index $j$ is rearranged to be in order from the smallest to the greatest $z_{i,j}$. Since the mean of $z_{i,j}$ is zero, the median immediately indicates the difference median – mean. The median is less sensitive to outliers than the mean.

Figure A1 (b) shows a histogram of the median, or median – mean, of the strip LCs in each detector. The dotted line represents $[z]_i$ calculated from the absolute value of the strip LCs, and the italic *n* denotes the sample numbers with $[z]_i < 0$ and $[z]_i \geq 0$ like in Fig. A1 (a). The histogram is biased to negative $[z]_i$. This demonstrates that a substantial number of detectors have large outliers in the strip LCs, as expected from Fig. A1 (a). On the other hand, the distribution of $[z]_i$ calculated from logarithmic LCs (solid line and normal n) shows a comparable number of samples in $[z]_i < 0$ and $[z]_i \geq 0$.

These results demonstrate that the log-mean of the 8 strips is more appropriate than simple mean of the absolute LC for use as a representative value of each detector. This is one of the reasons we used 8-strip log-mean of the LC in sections 3 and 4.

## A2.2. Distribution profile in each strip position

*Location of Figure A2.*

The normalization introduced in subsection A2.2 allows us to investigate dependences of the performance parameters on the strip positions A to H free from the variations between detectors. Figure A2 shows a histogram of $z_{i,j}$ calculated from the logarithmic LC in each strip position. We can see that the central strips, D and E, have relatively small deviation of the distribution, while the edge strips, A and H, have significantly broad distributions.

In the case where the surface current is dominant in the LC, edge strips are expected to have relatively high LCs comparing to the central strips, because their electrodes have long edges in contact with the detector perimeter, or the guard ring. The absence of such a tendency in Fig. A2 demonstrates that the bulk current is dominant compared to the surface current, which is consistent with the results of the multivariate analyses in the main text.

*Location of Figure A3.*

Figure A3 shows a cross-section of a sample detector after the Li-drift process and copper-staining. The edge strip is close to the edge of the drifted area, which approximately corresponds to the guard-ring area. It can be expected that the thickness of the undrifted layer under the edge strip easily becomes non-uniform or thick due to a slight non-uniformity of the drifted layer. This is thought to be the reason that the histograms in Fig. A2 show relatively large deviations in the edge strips, because the undrifted layer works to suppress the bulk current. We note, however, that the non-uniformity is at an acceptable level as demonstrated in the main text. This analysis normalizes the inter-strip variation in each detector, enhancing even small fluctuations in good performance detectors.

### A3. Strip capacitances

*Location of Figure A4 and A5.*

We calculate $z_{i,j}$ using the strip capacitance [pF] as $x_{i,j}$. Figure A4(a) shows a histogram of $z_{i,j}$. Even if certain operations on the measured value, such as logarithmic conversion, are not performed, the histogram is symmetric and well centered around the zero on the horizontal axis. Figure A4 (b) shows a histogram of the difference, median – mean of $z_{i,j}$ in each detector. The number of detectors (n) is larger in $[z]_i < 0$ than $[z]_i \geq 0$, but the deviation in the histogram is smaller than that of Fig. A1 (b).

These results demonstrate the validity of using a simple mean of 8-strip capacitances as a representative value of each detector.

Figure A5 displays a histogram of the normalized capacitance at each strip position. The edge strips A and H show relatively broad distributions, like in Fig. A2. The same interpretations as the strip LC can be given for this tendency.

## Acknowledgments

M. Kozai is supported by the JSPS KAKENHI grants JP17K14313 and JP20K14505. H. Fuke is supported by the JSPS KAKENHI grants JP17H01136 and JP19H05198 and from the Mitsubishi Foundation. K. Perez and M. Xiao are supported by the Heising-Simons Foundation. F. Rogers is supported by the NSF Graduate Research Fellowship grant 1122374. Y. Shimizu is supported by the JSPS KAKENHI grant JP20K04002 and Sumitomo Foundation. This work is partially supported by the JAXA/ISAS Small Science Program in Japan and the NASA APRA program through grants NNX17AB44G and NNX17AB46G in US.

We thank SUMCO Corporation and Shimadzu Corporation for their cooperation in our detector development and mass production.

## References

[1] P. v. Doetinchem, K. Perez, T. Aramaki, S. Baker, S. Barwick, R. Bird, M. Boezio, S.E. Boggs, M. Cui, A. Datta, F. Donato, C. Evoli, L. Fabris, L. Fabbietti, E.F. Bueno, N. Fornengo, H. Fuke, C. Gerrity, D.G. Coral, C. Hailey, D. Hooper, M. Kachelriess, M. Korsmeier, M. Kozai, R. Lea, N. Li, A. Lowell, M. Manghisoni, I. V. Moskalenko, R. Munini, M. Naskret, T. Nelson, K.C.Y. Ng, F. Nozzoli, A. Oliva, R.A. Ong, G. Osteria, T. Pierog, V. Poulin, S. Profumo, T. Pöschl, S. Quinn, V. Re, F. Rogers, J. Ryan, N. Saffold, K. Sakai, P. Salati, S. Schael, L. Serksnyte, A. Shukla, A. Stoessl, J. Tjemsland, E. Vannuccini, M. Vecchi, M.W. Winkler, D. Wright, M. Xiao, W. Xu, T. Yoshida, G. Zampa, P. Zuccon, Cosmic-ray antinuclei as messengers of new physics: Status and outlook for the new decade, J. Cosmol. Astropart. Phys. 2020 (2020) 1–42. https://doi.org/10.1088/1475-7516/2020/08/035.

[2] M. Kozai and the GAPS collaboration, The GAPS experiment - A search for cosmic-ray antinuclei from dark matter, J. Phys. Conf. Ser. 1468 (2020). https://doi.org/10.1088/1742-6596/1468/1/012049.

[3] H. Fuke, T. Aramaki, S. Boggs, W.W. Craig, P. v. Doetinchem, L. Fabris, C.J. Hailey, F. Gahbauer, T. Gordon, C. Kato, A. Kawachi, T. Koike, M. Kozai, N. Madden, S.A.I. Mognet, K. Mori, K. Munakata, S. Okazaki, R.A. Ong, K. Perez, K. Sakimoto, Y. Shimizu, N. Yamada, A. Yoshida, T. Yoshida, K.P. Ziock, J. Zweerink, Present status and future plans of GAPS antiproton and antideuteron measurement for indirect dark matter search, JPS Conf. Proc. 18 (2017) 1–6. https://doi.org/10.7566/jpscp.18.011003.

[4] T. Aramaki, C.J. Hailey, S.E. Boggs, P. v. Doetinchem, H. Fuke, S.I. Mognet, R.A. Ong, K. Perez, J. Zweerink, Antideuteron sensitivity for the GAPS experiment, Astropart. Phys. 74 (2016) 6–13. https://doi.org/10.1016/j.astropartphys.2015.09.001.

[5] E.M. Pell, Ion drift in an n-p junction, J. Appl. Phys. 31 (1960) 291–302. https://doi.org/10.1063/1.1735561.

[6] F.S. Goulding, Semiconductor detectors for nuclear spectrometry, I, Nucl. Instruments Methods. 43 (1966) 1–54.

[7] W.R. Umverstty, Fabrication methods for lithium drifted surface barrier silicon detectors, Nucl. Instruments Methods. 40 (1966) 330–336.

[8]  E. Belcarz, J. Chwaszczewska, M. Slapa, M. Szymczak, J. Tys, Surface barrier lithium drifted silicon detector with evaporated guard ring, Nucl. Instruments Methods. 77 (1970) 21–28.

[9]  C.E. Lyman, D.B. Williams, J.I. Goldstein, X-ray detectors and spectrometers, Ultramicroscopy. 28 (1989) 137–149. https://doi.org/10.1016/0304-3991(89)90286-6.

[10] C.S. Rossington, J.T. Walton, J.M. Jaklevic, Si(Li) detectors with thin dead layers for low energy X-ray detection, IEEE Trans. Nucl. Sci. 38 (1991) 239–243.

[11] C.E. Cox, D.A. Fischer, W.G. Schwarz, Y. Song, Improvement in the low energy collection efficiency of Si(Li) X-ray detectors, Nucl. Instruments Methods Phys. Res. Sect. B Beam Interact. with Mater. Atoms. 241 (2005) 436–440. https://doi.org/10.1016/j.nimb.2005.07.091.

[12] T. Miyachi, S. Ohkawa, T. Emura, M. Nishimura, O. Nitoh, K. Takahashi, S. Kitamura, Y. Kim, T. Abe, H. Matsuzawa, A thick and large active area Si ( Li ) detector, Jpn. J. Appl. Phys. 27 (1988) 307–310.

[13] T. Miyachi, S. Ohkawa, H. Matsuzawa, T. Otogawa, N. Kobayashi, Y. Ikeda, H. Onabe, Development of lithium-drifted silicon detectors using an automatic lithium-ion drift apparatus, Jpn. J. Appl. Phys. 33 (1994) 4115. https://doi.org/10.1143/JJAP.33.4115.

[14] T. Kashiwagi, J. Kikuchi, T. Nakasugi, A. Nakamoto, A fabrication method and problems of large area Si(Li) detector, Report, Waseda Res. Inst. Sci. Eng. 128 (1990) 66–80.

[15] T. Kohno, K. Munakata, T. Imai, A. Yoneda, C. Kato, M. Matsuoka, T. Doke, J. Kikuchi, T. Kashiwagi, K. Nishijima, N. Hasebe, H.J. Crawford, Accelerator beam experiments of a prototype cosmic ray heavy ion telescope, J. Phys. Soc. Japan. 60 (1991) 3967–3975.

[16] H. Onabe, H. Kume, M. Obinata, T. Kashiwagi, Development of Si(Li) detectors for charged particles spectrometer, Ioniz. Radiat. 28 (2002) 159–167.

[17] T. Takashima, T. Kashnwagi, S. Okuno, K. Mori, H. Onabe, The development of the high energy particle detector onboard the SELENE spacecraft, IEEE Nucl. Sci. Symp. Conf. Rec. 1 (2004) 181–185. https://doi.org/10.1109/nssmic.2004.1462177.

[18] M. Kozai, H. Fuke, M. Yamada, K. Perez, T. Erjavec, C.J. Hailey, N. Madden, F. Rogers, N. Saffold, D. Seyler, Y. Shimizu, K. Tokuda, M. Xiao, Developing a mass-production model of large-area Si(Li) detectors with high operating temperatures, Nucl. Instruments Methods Phys. Res. Sect. A Accel. Spectrometers, Detect. Assoc. Equip. 947 (2019) 162695. https://doi.org/10.1016/j.nima.2019.162695.

[19] K. Perez, T. Aramaki, C.J. Hailey, R. Carr, T. Erjavec, H. Fuke, A. Garvin, C. Harper, G. Kewley, N. Madden, S. Mechbal, F. Rogers, N. Saffold, G. Tajiri, K. Tokuda, J. Williams, M. Yamada, Fabrication of low-cost , large-area prototype Si ( Li ) detectors for the GAPS experiment, Nucl. Instruments Methods Phys. Res. Sect. A Accel. Spectrometers, Detect. Assoc. Equip. 905 (2018) 12–21. https://doi.org/10.1016/j.nima.2018.07.024.

[20] F. Rogers, M. Xiao, K. Perez, T. Erjavec, L. Fabris, H. Fuke, J. Hailey, M. Kozai, A. Lowell, N. Madden, S. Mcbride, V. Re, E. Riceputi, N. Saffold, Y. Shimizu, Large-area Si(Li) detectors for X-ray spectrometry and particle tracking in the GAPS experiment, J. Instrum. 14 (2019) 1–16. https://doi.org/10.1088/1748-0221/14/10/P10009.

[21] N. Saffold, F. Rogers, M. Xiao, R. Bhatt, T. Erjavec, H. Fuke, C.J. Hailey, M. Kozai, D. Kraych, E. Martinez, C. Melo-Carrillo, K. Perez, C. Rodriguez, Y. Shimizu, B. Smallshaw, Passivation of Si(Li) detectors operated above cryogenic temperatures for space-based applications, Nucl. Instruments Methods Phys. Res. Sect. A Accel. Spectrometers, Detect. Assoc. Equip. 997 (2021) 165015. https://doi.org/10.1016/j.nima.2020.165015.

[22] S. Okazaki, H. Fuke, H. Ogawa, Y. Miyazaki, K. Takahashi, N. Yamada, Meter-scale multi-loop capillary heat pipe, Appl. Therm. Eng. 141 (2018) 20–28. https://doi.org/10.1016/j.applthermaleng.2018.05.116.

[23] H. Fuke, S. Okazaki, H. Ogawa, Y. Miyazaki, Balloon flight demonstration of an oscillating heat pipe, J. Astron. Instrum. 6 (2017) 1–10. https://doi.org/10.1142/S2251171717400062.

[24] S. Okazaki, H. Fuke, T. Inoue, A. Kawachi, D. Matsumoto, Y. Miyazaki, H. Nagai, T. Nonomura, H. Ogawa, N. Yamada, Conceptual design study of oscillating heat pipe system for GAPS, Proc. 30th Intl. Symp. Sp. Technol. Sci. (2015) 1–5.

[25] H. Fuke, T. Abe, T. Daimaru, T. Inoue, A. Kawachi, H. Kawai, Y. Masuyama, H. Matsumiya, D. Matsumoto, Y. Miyazaki, J. Mori, H. Nagai, T. Nonomura, H. Ogawa, S. Okazaki, T. Okubo, S. Ozaki, D. Sato, K. Shimizu, K. Takahashi, S. Takahashi, N. Yamada, T. Yoshida, Development of a cooling system for GAPS using oscillating heat pipe, Trans. Japan Soc. Aeronaut. Sp. Sci. Aerosp. Technol. Japan. 14 (2016) Pi_17-Pi_26. https://doi.org/10.2322/tastj.14.pi_17.

[26] S. Hirao, K. Nakata, A comprehensive yield monitoring system for high-mix low-volume semiconductor, 34th Annu. Conf. Japanese Soc. Artif. Intell. (2020) 1–4. https://doi.org/https://doi.org/10.11517/pjsai.JSAI2020.0_4O3GS1303.

[27] T. Hidetaka, H. Shirai, M. Terabe, K. Hashimoto, A. Shinohara, Analysis methodology for semiconductor yield by data mining, IEEJ Trans. Ind. Appl. 129 (2009) 1201–1211. https://doi.org/10.1541/ieejias.129.1201.

[28] C.F. Chien, W.C. Wang, J.C. Cheng, Data mining for yield enhancement in semiconductor manufacturing and an empirical study, Expert Syst. Appl. 33 (2007) 192–198. https://doi.org/10.1016/j.eswa.2006.04.014.

[29] K.R. Skinner, D.C. Montgomery, G.C. Runger, J.W. Fowler, D.R. McCarville, T.R. Rhoads, J.D. Stanley, Multivariate statistical methods for modeling and analysis of wafer probe test data, IEEE Trans. Semicond. Manuf. 15 (2002) 523–530. https://doi.org/10.1109/TSM.2002.804901.

[30] V. Raghavan, Application of decision trees for integrated circuit yield improvement, IEEE Int. Symp. Semicond. Manuf. Conf. Proc. (2002) 262–265. https://doi.org/10.1109/asmc.2002.1001615.

[31] A. Fong, J.T. Walton, E.E. Haller, H.A. Sommer, J. Guldberg, Characterization of large diameter silicon by low-bias charge collection analysis in Si(Li) pin diodes, Nucl. Instruments Methods Phys. Res. 199 (1982) 623–630. https://doi.org/10.1016/0167-5087(82)90164-8.

[32] P.G. Litovchenko, W. Wahl, D. Bisello, R. Rando, A.P. Litovchenko, V.F. Lastovetsky, L.I. Barabash, T.I. Kibkalo, L.A. Polivtsev, J.I. Kolevatov, V.P. Semenov, L.A. Trykov, J. Wyss, Silicon detectors for γ-ray and β-spectroscopy, Nucl. Instruments Methods Phys. Res. Sect. A Accel. Spectrometers, Detect.

Assoc. Equip. 512 (2003) 408–411. https://doi.org/10.1016/S0168-9002(03)01919-3.

[33]  P.J. Whoriskey, Two chemical stains for marking p-n junctions in silicon, J. Appl. Phys. 29 (1958) 867–868. https://doi.org/10.1063/1.1723306.

[34]  H. Kume, H. Onabe, M. Obinata, T. Kashiwagi, Evaluation of Si(Li) detectors by a combination of the copper plating method and x-ray analytical microscopy, IEEE Trans. Nucl. Sci. 48 (2001) 1012–1015. https://doi.org/10.1109/23.958715.

[35]  F. Rogers, M. Xiao, K.M. Perez, S. Boggs, T. Erjavec, L. Fabris, H. Fuke, C.J. Hailey, M. Kozai, A. Lowell, N. Madden, M. Manghisoni, S. McBride, V. Re, E. Riceputi, N. Saffold, Y. Shimizu, Large-area Si(Li) detectors for X-ray spectrometry and particle tracking in the GAPS experiment, J. Instrum. 14 (2019). https://doi.org/10.1088/1748-0221/14/10/P10009.

[36]  R Core Team, R: A language and environment for statistical computing, R Found. Stat. Comput. Vienna, Austria. (n.d.). https://www.r-project.org/.

[37]  H. McDuff, M.R. Hoeferkamp, S. Seidel, R. Wang, C.J. Kenney, J. Hasi, S.I. Parker, The effect of humidity on reverse breakdown in 3D silicon sensors, Nucl. Instruments Methods Phys. Res. Sect. A Accel. Spectrometers, Detect. Assoc. Equip. 785 (2015) 1–4. https://doi.org/10.1016/j.nima.2015.02.056.

# Figures

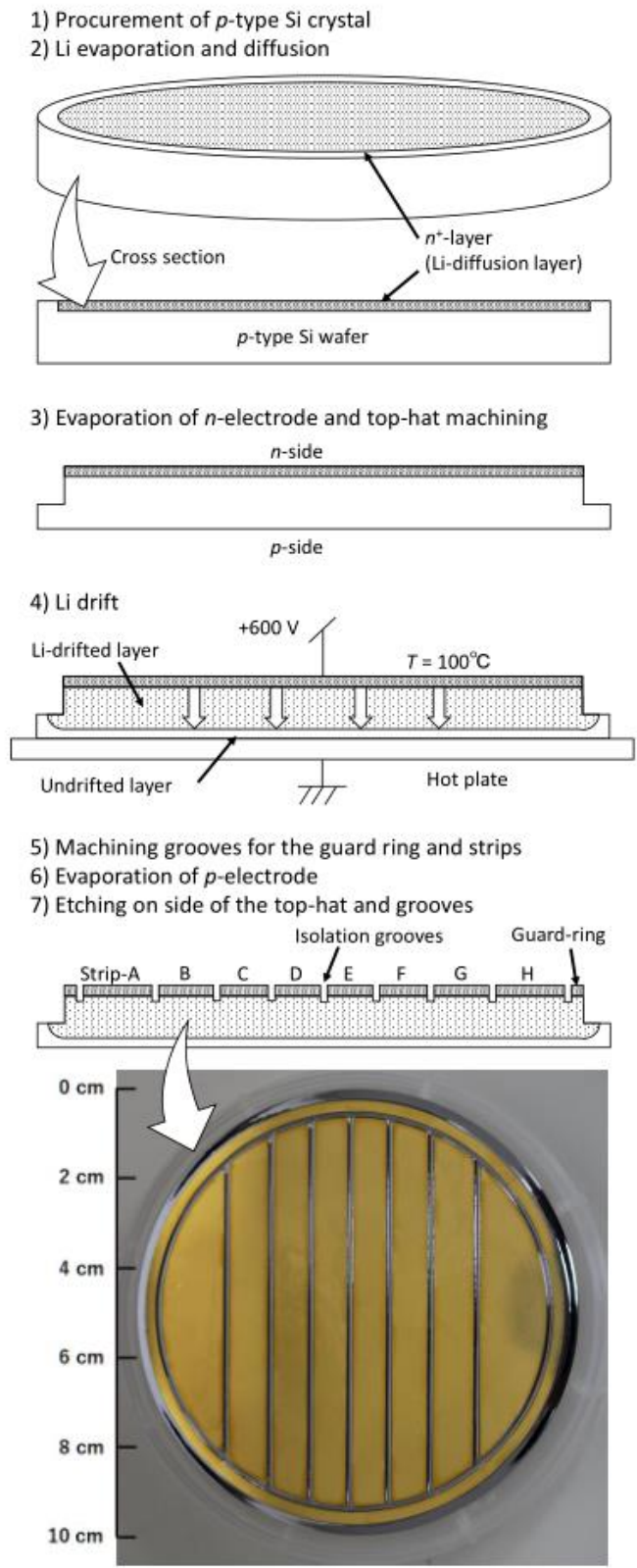

Figure 1. Fabrication process of the GAPS Si(Li) detector.

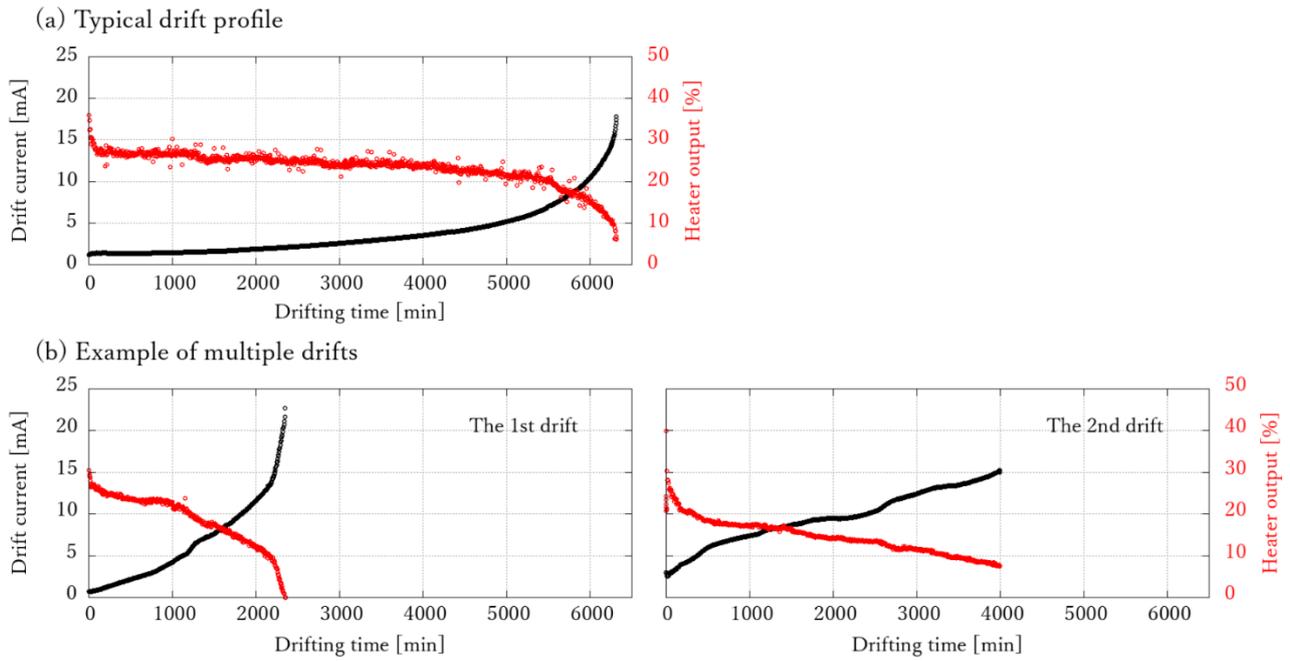

Figure 2. Examples of the Li-drift profiles. (a) Typical drift profile. (b) Example of multiple drifts. Black points (left vertical axis) indicate the drift current [mA] while red points (right vertical axis) indicate the heater output in arbitrary unit [%].

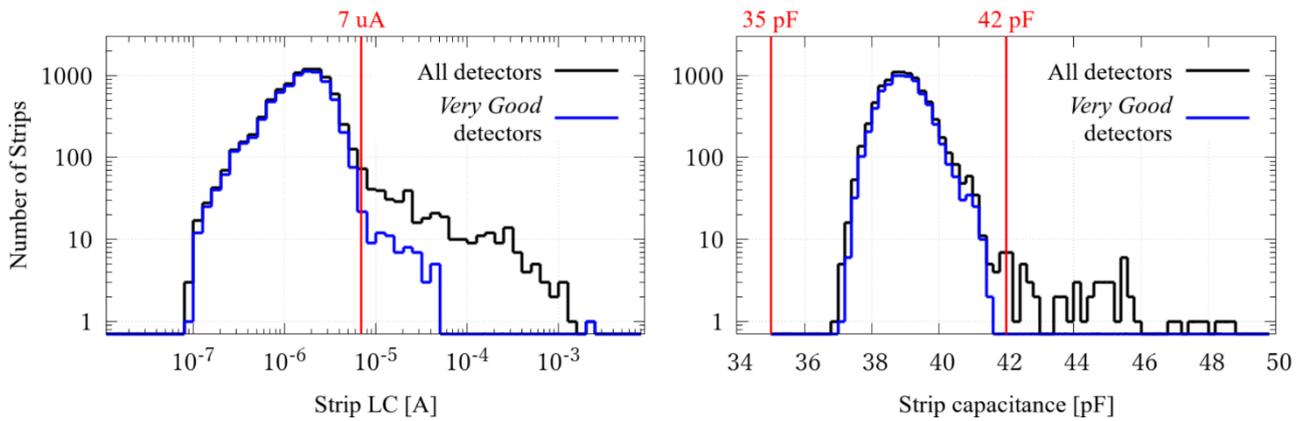

Figure 3. Histograms of the strip LC (left) and capacitance (right) at RT for all detectors (black line) and *Very Good* detectors (blue line). The red vertical lines represent values in the *Very Good* criteria.

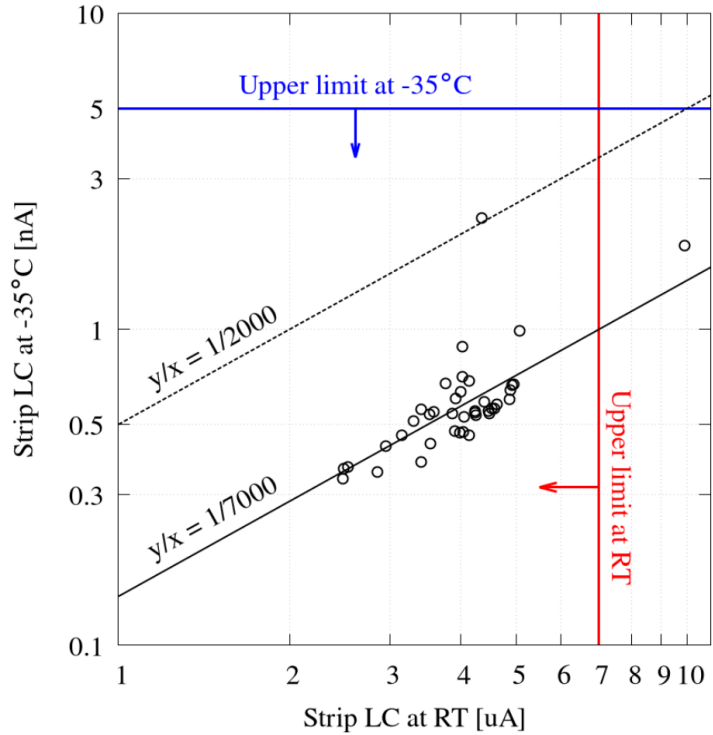

Figure 4. Strip LCs at RT (horizontal axis) and −35°C (vertical axis) measured for 40 strips of 5 detectors among the mass-produced detectors. The blue horizontal line indicates the upper limit at −35°C, while the red vertical line indicates the upper limit in the *Very Good* criteria.

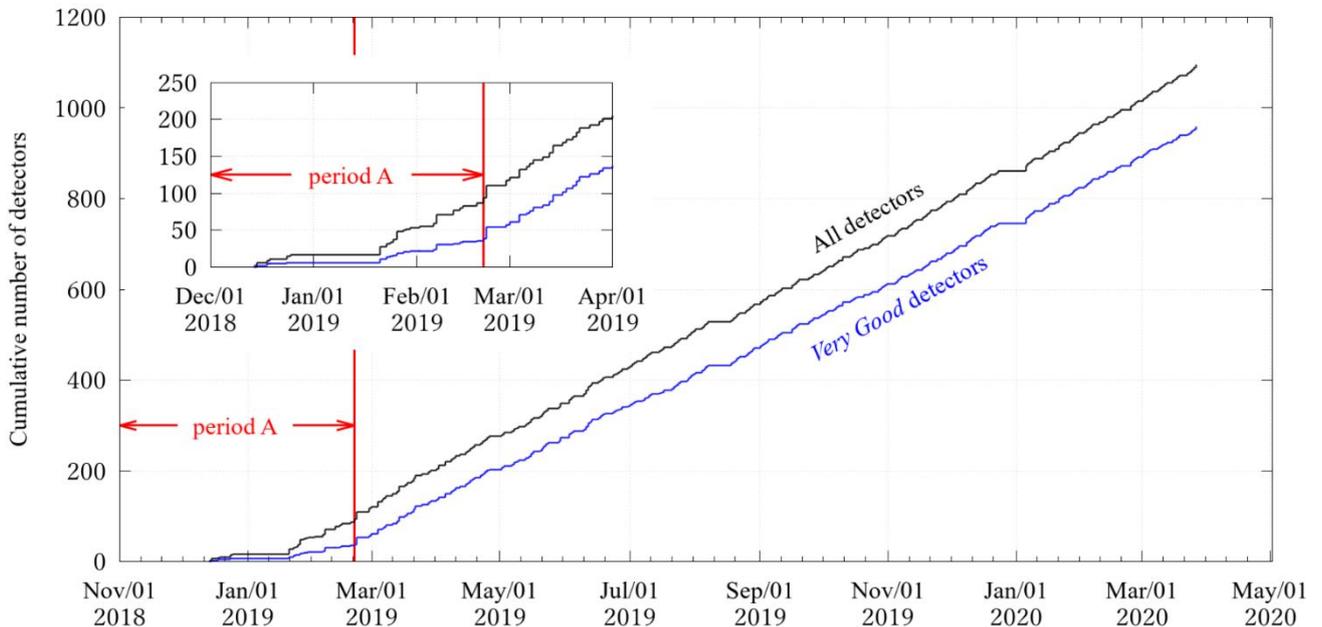

Figure 5. Cumulative number of detectors during mass production. The black curve shows the total number of detectors. The number of *Very Good* detectors is indicated by a blue curve. Red arrows indicate the period (period A) until the middle of February 2019 when the upper limit on the drift time is adopted. The upper left figure shows an enlargement of the initial stage of the mass production including period A.

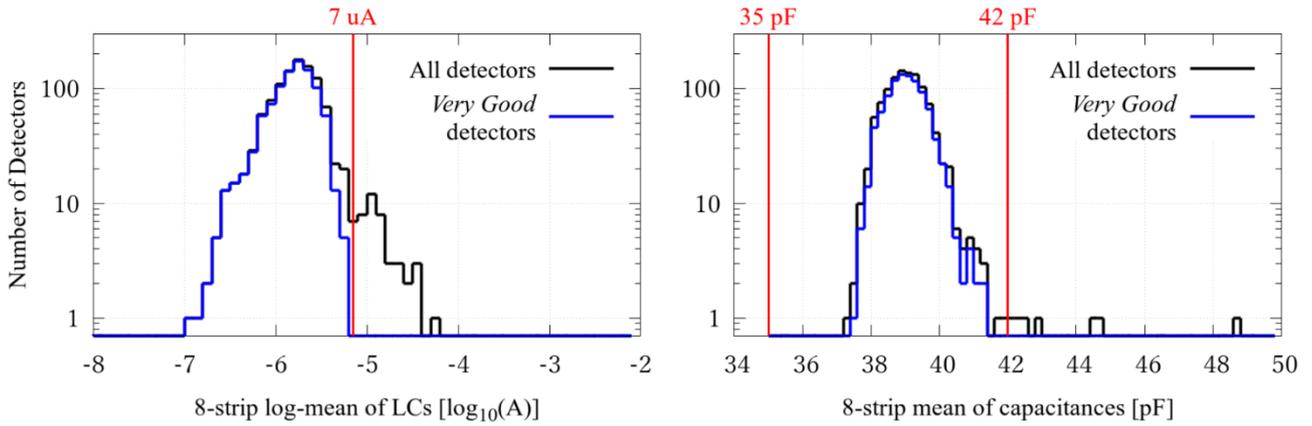

Figure 6. Distributions of (left) 8-strip log-mean of LCs ($\langle \log_{10} I \rangle_i$) and (right) 8-strip mean of capacitances ($\langle C \rangle_i$) in each detector. The black histograms show the distributions for all detectors while the blue histograms show distributions of the *Very Good* detectors. Red vertical lines represent values in the *Very Good* criteria.

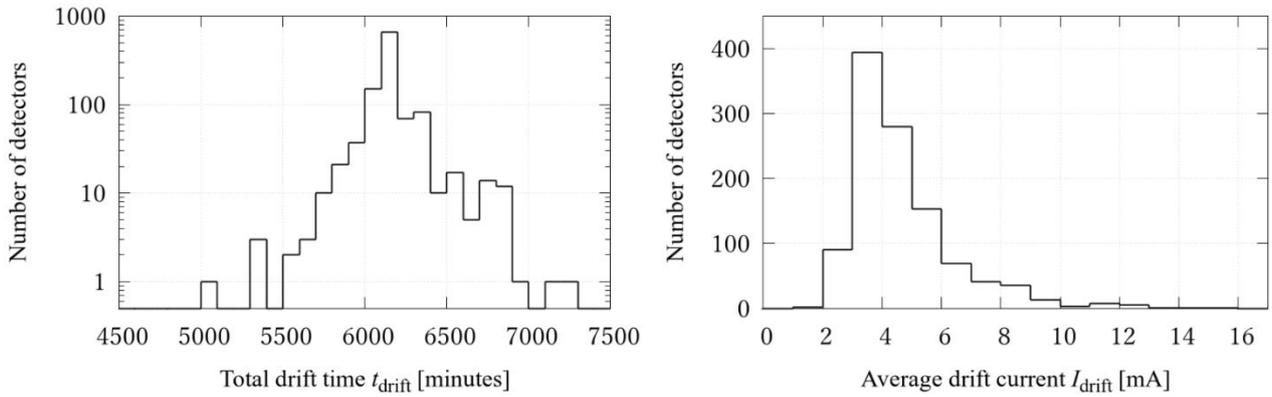

Figure 7. Histograms of the total drift time $t_{\text{drift}}$ and average drift current $I_{\text{drift}}$. Note that the vertical axis is displayed in logarithmic scale in the left histogram ($t_{\text{drift}}$).

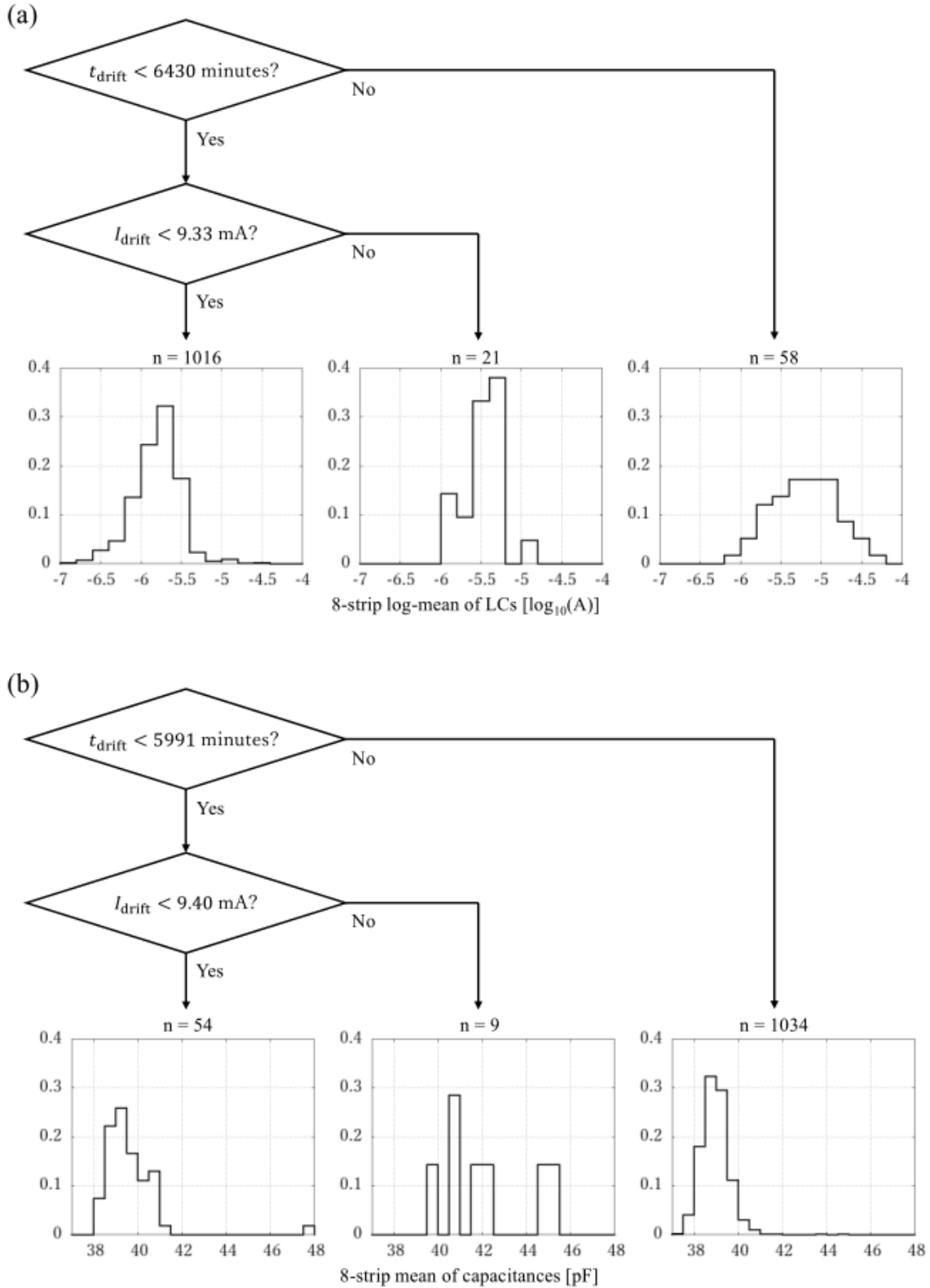

Figure 8. Results of the regression tree analyses for (a) 8-strip log-mean of LC and (b) 8-strip mean of capacitances. The flow chart describes the derived discriminants, and the histograms below display the distributions of the objective variables in each branch of the tree. In each histogram, the total number of detectors is normalized to 1

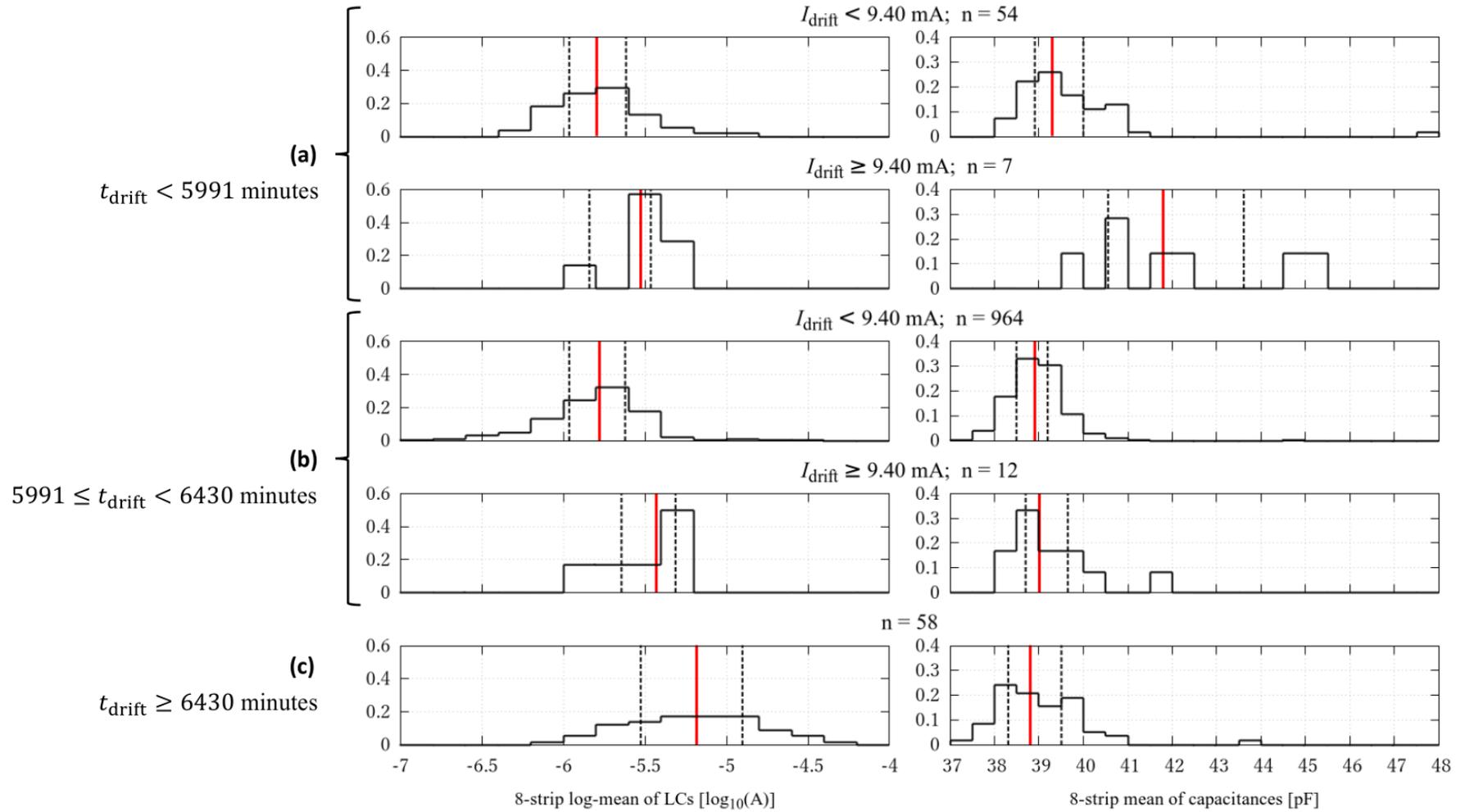

Figure 9. Histograms of the LC and capacitance in 5 groups classified by the total drift time $t_{drift}$ and average drift current $I_{drift}$ based on the results of the regression tree analyses. The median of each histogram is represented by a red vertical line, and the first and third quartiles are shown by dotted vertical lines. In each histogram, the total number of detectors is normalized to 1.

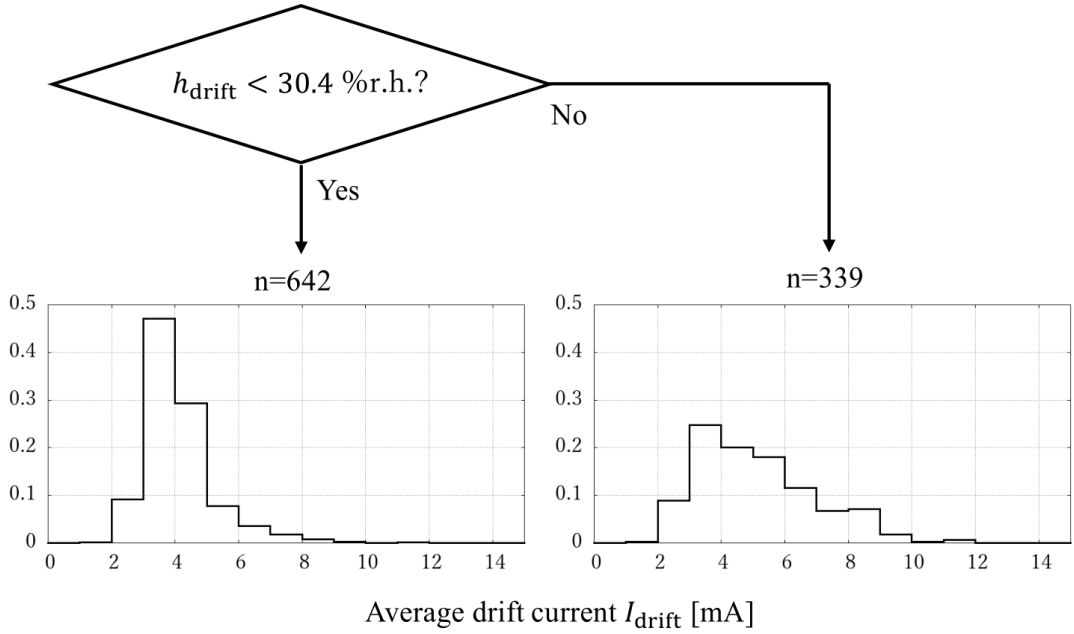

Figure 10. Regression tree for the average drift current $I_{\text{drift}}$. Other descriptions are the same as Fig. 8.

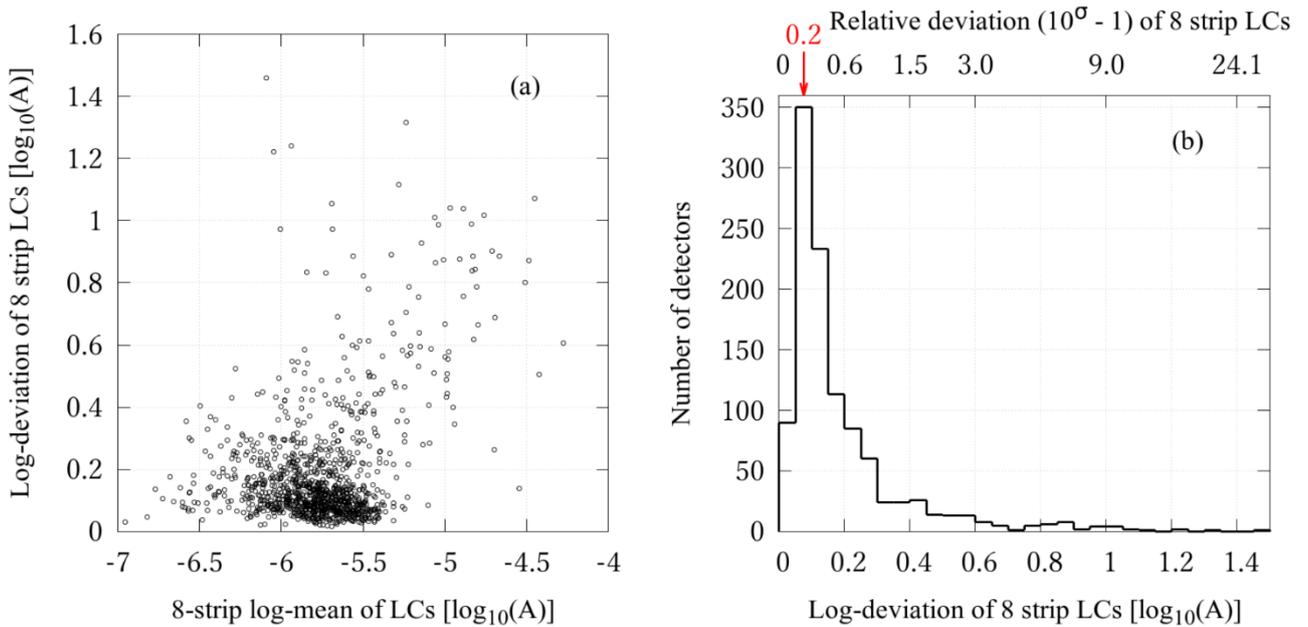

Figure 11. (a) Scatter plot between the 8-strip log-mean ($\langle \log_{10} I \rangle_i$) and log-deviation ($\sigma_i(\log_{10} I)$) of LCs in each detector. (b) Histogram of the 8-strip log-deviation of the LCs. The relative deviation of the strip LCs in each detector, corresponding to each log-deviation value, is displayed on the upper horizontal axis in panel (b). The red arrow indicates the relative deviation at the mode of the histogram.

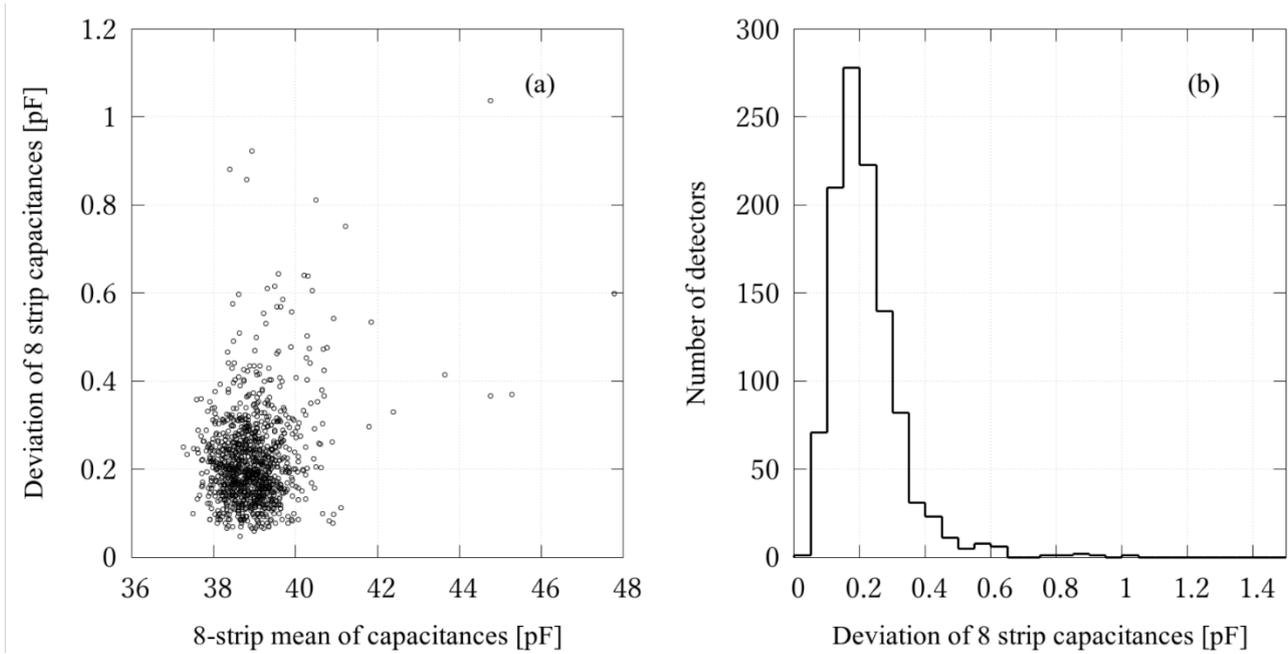

Figure 12. (a) Scatter plot between the 8-strip mean ($\langle C \rangle_i$) and deviation ($\sigma_i(C)$) of the capacitances in each detector. (b) Histogram of $\sigma_i(C)$.

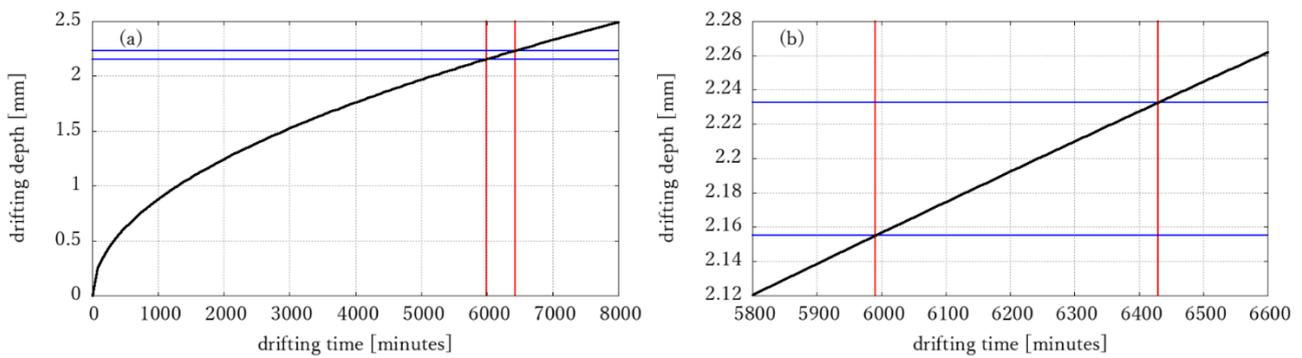

Figure 13. (a) Relationship between drift time $t$ and drift depth $W$ based on the theoretical equation (1). Lower and upper limits of the total drift time (5991 min and 6430 min) obtained by the regression tree analyses are shown by red vertical lines. The blue horizontal lines are drift depths each calculated from drift time by equation (1). Panel (b) is an enlargement of the range between the red lines.

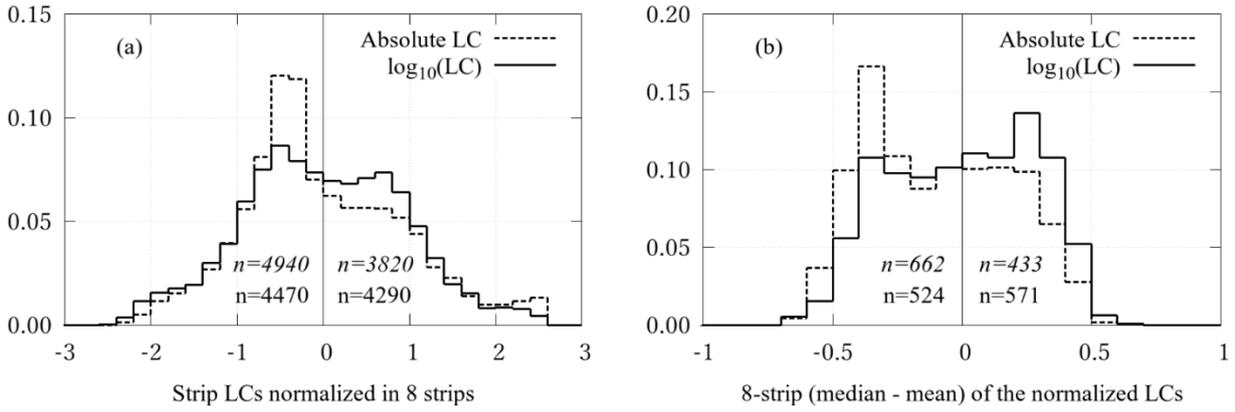

Figure A1. (a) Distribution of the strip LC normalized in the 8 strips in each detector. The dotted line displays a histogram of $z_{i,j}$ calculated by using the absolute LC [A] as $x_{i,j}$, while the solid line is deduced by using the logarithmic LC [$\log_{10}(A)$] as $x_{i,j}$. The total number of strips is normalized to 1. (b) Distribution of the difference, (median – mean) of 8 values of $z_{i,j}$ in each detector. The total number of detectors is normalized to 1.

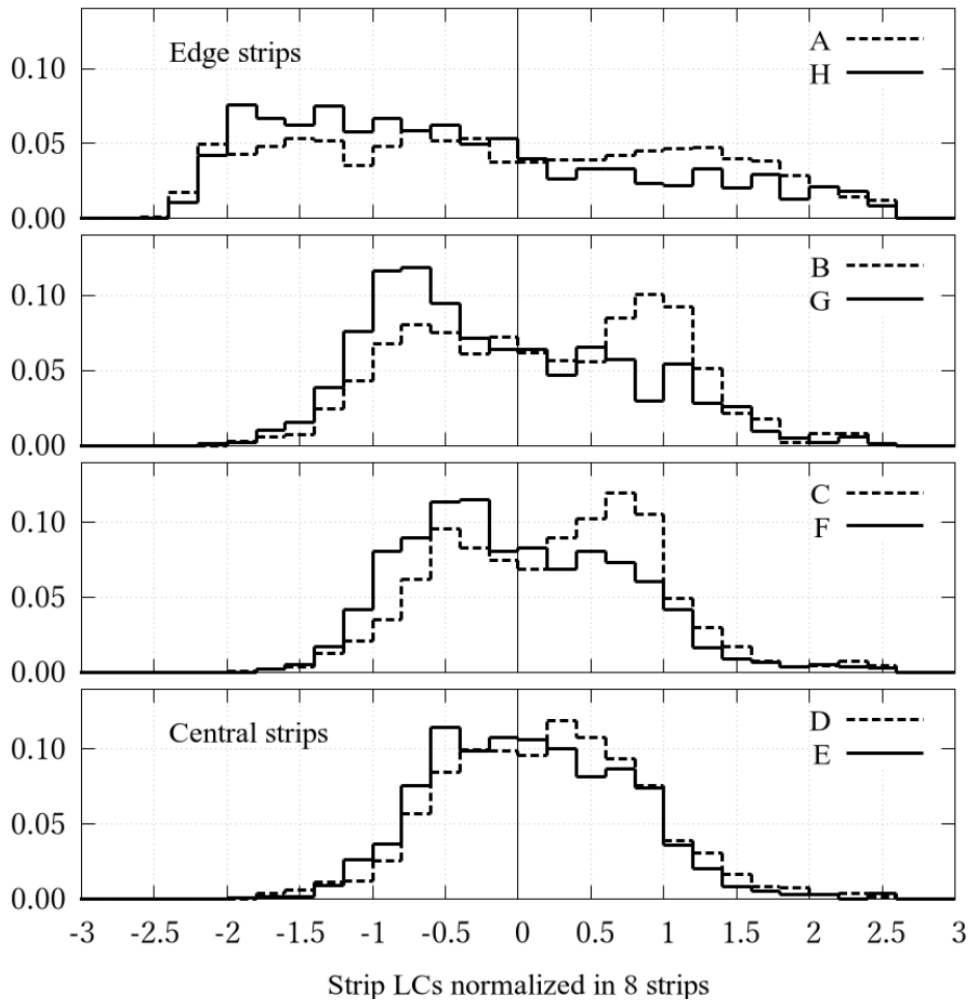

Figure A2. Histogram of the normalized logarithmic LC in each strip position (A to H). The top panel displays edge strips (A and H) and the bottom panel shows central strips (D and E). In each histogram, the total number of strips is normalized to 1.

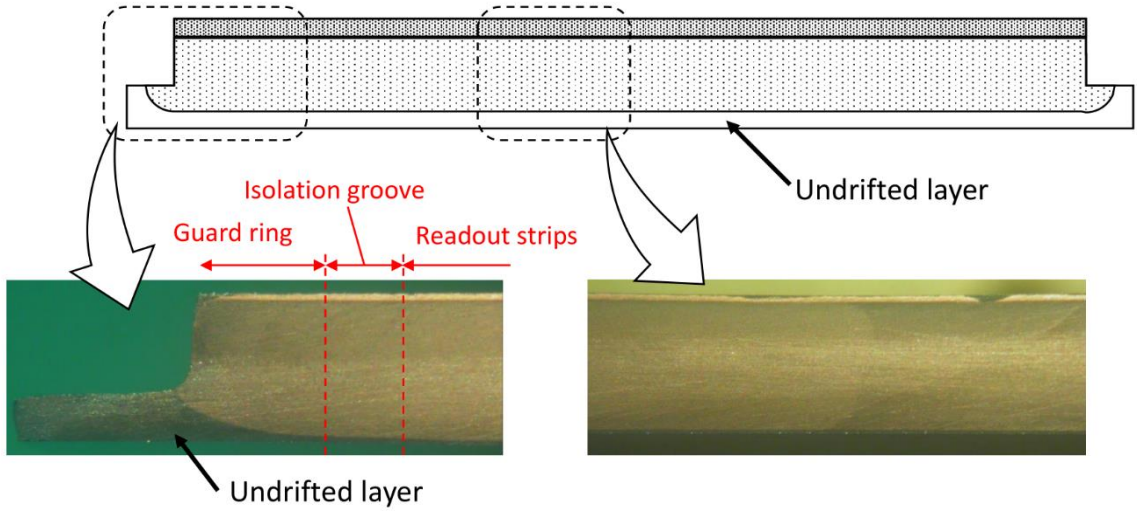

Figure A3. Cross-section of a sample detector after the Li-drift process and copper-staining. Areas of the guard ring, isolation groove, and readout strips in the finished detector are displayed in the left picture. (Modification from Fig. 5 in [18].)

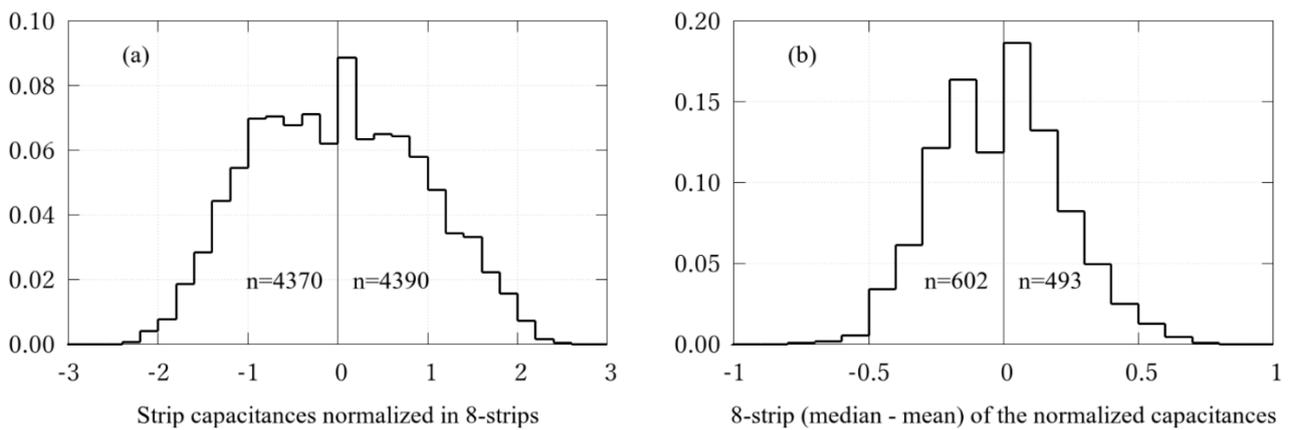

Figure A4. (a) Distribution of the strip capacitance normalized in the 8 strips in each detector. (b) Distribution of the difference, (median – mean) of 8 values of $z_{i,j}$ in each detector. Each histogram is normalized in the same manner as Fig. A1.

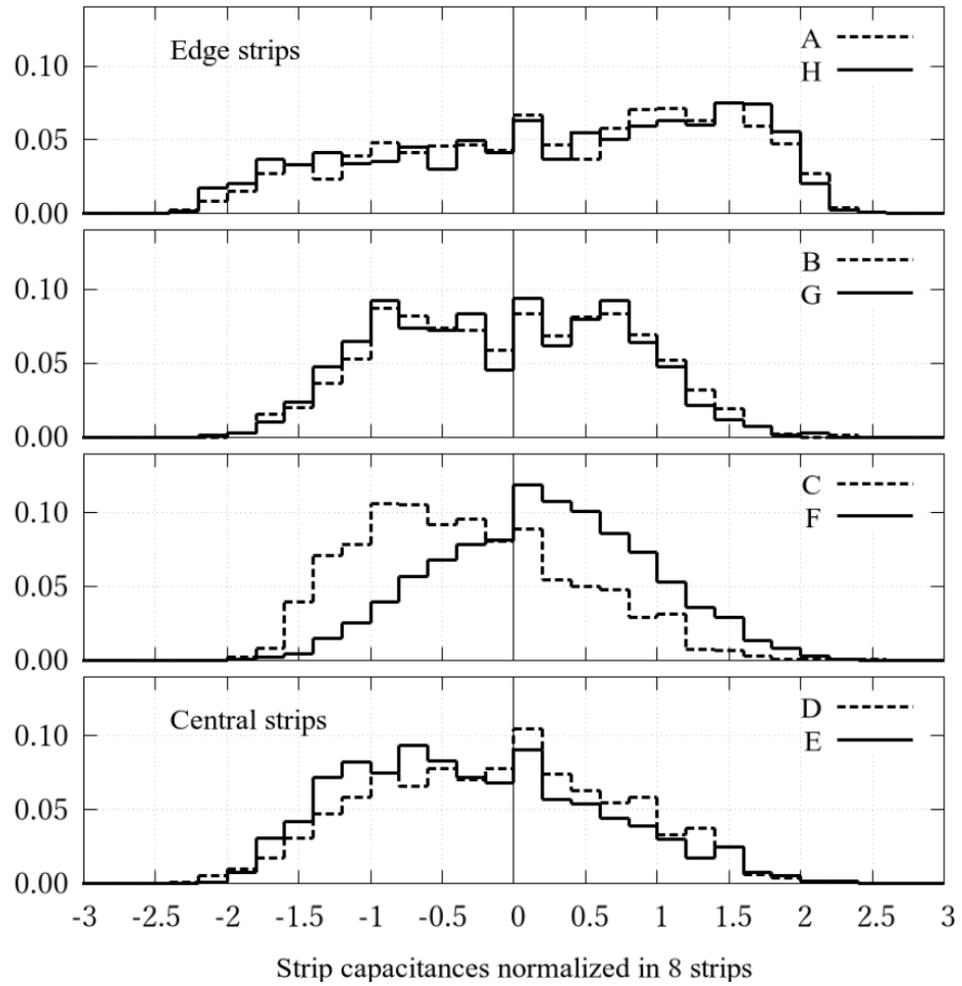

Figure A5. Histogram of the normalized capacitance in each strip position (A to H). The top panel shows edge strips (A and H) and the bottom panel shows central strips (D and E). Each histogram is normalized in the same manner as Fig. A2.